\documentclass[conference]{IEEEtran}
\IEEEoverridecommandlockouts
\usepackage{bbding}

\usepackage{cite}
\usepackage{amsmath,amssymb,amsfonts}
\usepackage{graphicx}
\usepackage{textcomp}
\usepackage{xcolor}
\usepackage{algorithm}
\usepackage{algpseudocode}
\def\BibTeX{{\rm B\kern-.05em{\sc i\kern-.025em b}\kern-.08em
    T\kern-.1667em\lower.7ex\hbox{E}\kern-.125emX}}
\begin{document}

\title{Human Biometric Signals Monitoring based on WiFi Channel State Information using Deep Learning\\
}




\author{Moyu Liu, \emph{Student Member}, Zihuai Lin, \emph{Senior Member}, Pei Xiao, \emph{Senior Member}, \tabularnewline
Wei Xiang, \emph{Senior Member} 

\thanks{This research is supported by Australian Research Council (ARC) Discovery
projects DP190101988.
Moyu Liu and Zihuai Lin are with the School of Electrical and Information Engineering,
The University of Sydney, Australia (e-mail: \{moyu.liu,zihuai.lin\}@sydney.edu.au).

Pei Xiao is with the Institute for Communication Systems (ICS), University of Surrey, UK (e-mail: p.xiao@surrey.ac.uk )

Wei Xiang is with the Cisco-La Trobe Centre for AI and Internet of Things,
La Trobe University, Australia (e-mail: W.Xiang@latrobe.edu.au).}
}

\maketitle

\begin{abstract}
In this paper, we first present a single-input, multiple-output convolutional neural network that can estimate both heart rate and respiration rate simultaneously by exploiting the underlying link between heart rate and respiration rate. The inputs to the neural network are the amplitude and phase of channel state information collected by a pair of WiFi devices. Our WiFi-based technique addresses privacy concerns and is adaptable to a variety of settings. This system’s overall accuracy for the heart and respiration rate estimation can reach $99.109\%$ and $98.581\%$, respectively. Furthermore, we developed and analyzed two deep learning-based neural network classification algorithms for categorizing four types of sleep stages: wake, rapid eye movement (REM) sleep, non-rapid eye movement (NREM) light sleep, and NREM deep sleep. This system’s overall classification accuracy can reach $95.925\%$.
\end{abstract}





\section{Introduction}

Health monitoring equipment is becoming ubiquitous with the ever increasing attention to public health. Most health monitoring instruments mainly detect human biometric signals, such as heart rate and respiratory rate. These indicators can reflect human primary health conditions, both mental and physical, and classify sleep stages. The classification of sleep stages, as well as the monitoring of respiration rate and heart rate, can aid the evaluation of health conditions and the diagnosis of disorders by doctors. Some health problems, such as depression, insomnia, obesity, and other diseases, may benefit from it.

With the advancement of wireless technologies, the era of smart health and smart medical care has arrived in society, which brought some clear benefits, particularly in the sphere of health care. According to statistics \cite{boric2002wireless}, more than 100 million people in the United States suffer from chronic conditions such as diabetes or heart disease. In addition, sleep problems are affecting more and more people. Sleep quality is closely related to some health problems, such as sleep apnea \cite{caples2005obstructive}, asthma \cite{braun2012bridging}, chronic insomnia \cite{farrell2017recognition}, diabetes \cite{tasali2008slow} and depression \cite{parish2009sleep}. Hence, monitoring human biometric signals, such as heart rate and respiratory rate, during sleep or daytime is essential for tracking human health and sleep status. According to a recent study of adult labor in the United States \cite{roth2009slow}, $16\%$ of participants had less than six hours of sleep throughout the workday. This has also heightened public awareness of the advancement of smart medical care and artificial intelligence biomedical technology in the field of sleep health.

With the rapid growth of the Internet of Things (IoT) and artificial intelligence, the problem of humans not being able to monitor their health and sleep quality at home has been alleviated to some extent. Researchers have developed many wearable health detection devices. However, due to the discomfort of long-term wear and privacy concerns, researchers began to design smart health care devices that allow for health monitoring to be carried out using common household appliances. For example, the basic physiological parameters of the human body, such as heart rate and respiration rate, may be monitored using WiFi at home. WiFi can be used for more than just connecting to the internet; it can also be utilized for basic health management. It can provide daily health status updates to the patient, such as heart rate and respiration, among other things. It can also send an online alert if the patient has an emergency. It may also keep track of the quality of the patient's sleep.

In this paper, a WiFi-based method combined with convolutional neural networks (CNN) to predict heart rate and respiratory rate is proposed. We collect the amplitude and phase of the channel state information (CSI) to detect heart rate and respiration rate via a pair of WiFi devices. The WiFi system is based on the 802.11n standard and uses a 56-carrier orthogonal frequency division multiplexing scheme. This method can be more precise than using only one source of frequency spectrum or time domain information because these two types of information are complimentary \cite{zeng2018fullbreathe}, making it more comprehensive in capturing information.
In addition, we also design and compare two neural network classification approaches using CSI data for categorizing four types of sleep stages, including wake, rapid eye movement (REM) sleep, non-rapid eye movement (NREM) light sleep, and NREM deep sleep. When collecting CSI data during sleep, the data includes not only heartbeat and respiratory statistics, but also information on human body movements.

The main contributions of this paper can be summarised as follows:

\begin{itemize}
\item To estimate the heart rate and respiration rate simultaneously, we first propose a CNN with a single input and two outputs, namely, the Heart Rate and Respiration Rate Network (H3RN). The CSI's intrinsic connection between heart rate and respiration rate is used in this network. This strategy, rather than the typical sophisticated feature-selection methods, reduces computational complexity while simultaneously improving accuracy.

\item We propose a WiFi sleep strategy that uses the heart and respiration rates to determine four different sleep states. This differs from existing research works, which rely on extracted CSI features for detection. To the best of our knowledge, this is the first time to use WiFi device to track sleep stages by estimating the heart and respiration rates.

\item We develop two neural networks, namely the WiFi sleep-stage neural network (W2SN) and the cardiopulmonary coupling (CPC) Neural Network, for sleep stage classification. It is demonstrated that that the W2SN outperforms the CPC network because besides the heart and respiration rates, the CSI captured by W2SN also contains human body movement information, which is not included in the CPC signal. The W2SN can accurately categorize four sleep phases and has a classification accuracy of  95.925$\%$. 

\item The main differences between our work and some of existing research works are summarized in Table ~\ref{table:ff}.

\begin{table*}\caption{Comparison between some of existing research works and our work}\label{table:ff}
\centering
{\centering}
\begin{tabular}{|c|c|c|c|c|c|c|}
\hline
Types & our paper &\cite{liu2018monitoring}& \cite{khamis2018cardiofi} &\cite{li2016mo}\\
\hline\hline
{Using CNN to jointly estimate heart and respiration rates} &\Checkmark &\XSolid &\XSolid &\XSolid \\ 
\hline
{Using amplitude and phase information for CSI} &\Checkmark &\XSolid &\XSolid &\XSolid \\  
\hline
{Detection of the heart and respiration rates under Different Postures} &\Checkmark &\Checkmark &\XSolid &\XSolid \\  
\hline
{Line-of-sight (LOS) Environment} &\Checkmark &\Checkmark &\Checkmark &\Checkmark \\ 

\hline
{Non-line-of-sight (NLOS) Environment} &\Checkmark &\XSolid &\XSolid &\XSolid \\ 
\hline
\end{tabular}
\end{table*}



\end{itemize}

The remainder of this paper is organized as follows. Related works are summarized in Section \uppercase\expandafter{\romannumeral2}. The architecture of the system is described in Section \uppercase\expandafter{\romannumeral3}. Section \uppercase\expandafter{\romannumeral4} describes the data processing and the neural network for the heart rate and respiration rate estimation. 
In Section \uppercase\expandafter{\romannumeral5}, we present the developed two neural networks based on the CPC and WiFi approaches to classify sleep stages. The experiment setup and performance are demonstrated in Section \uppercase\expandafter{\romannumeral6}. Finally, concluding remarks are drawn in Section \uppercase\expandafter{\romannumeral7}.

\section{Related Work}
Human biometric signals such as heart rate and respiratory rate are important indicators for monitoring sleep quality and physical health. There are two ways of monitoring human biometric signals during sleep: contact-based and non-contact-based.

\subsection{Contact-Based Methods} 
Polysomnography (PSG) \cite{sleep1998methods,van2011objective} is a contact-based method that is commonly used in clinical settings to monitor sleep quality. It is precise and complex, detecting a wide range of indicators via contact sensors, including body movement (detection via electrooculograms, EOG), respiration rate and heart rate (detection via electrocardiogram, ECG), brain activity (detection via EEG), and so on. EOG, for example, can distinguish the patient's sleep stage by measuring eye movement. Electromyography (EMG) can help determine sleep stage by monitoring the changes of muscle tone during sleep. PSG, which detects heart rate, respiration, brain waves, body movement, blood oxygen level, eye movement, abdominal movement, etc., is a multi-sensor approach with a large number of parameters. This approach necessitates a precise connection to different human body parts using various sensors, as well as technical professionals monitoring patients throughout the night.

Although the PSG-based method for measuring sleep stages is comprehensive and accurate, it has a number of limitations. Due to the high cost of the equipment, thus limited facility, patients need to spend a significant amount of waiting times, which may cause treatment to be delayed. Furthermore, this method is accompanied by some uncertain factors, resulting in inaccurate data monitoring. For example, an unfamiliar environment can cause insomnia, and emotional stress can affect the accuracy of recorded sleep quality data.

Photoplethysmography (PPG) can measure the blood volume changes in tissue based on optical sensors. It is a method for illuminating the skin with a light source and detecting the amount of light in the photodiode after being partly absorbed and scattered  in the tissue. This technique can be applied to detect heart rate and respiration rate \cite{alian2014photoplethysmography,nilsson2013respiration}or some sleep disorders \cite{karmakar2013detection,lazaro2012osas,haba2005obstructive}. For example, a respiration arousal detection model \cite{karmakar2013detection} has been developed by using the feature extraction from PPG.

Researchers have developed medical electronic equipment that allows people to obtain biometric information at any time by reducing the number of sensors. This equipment uses fewer physiological signals to track individual biological indicators. For instance, the PSYCHE system \cite{lanata2014complexity} with smartphone platform can monitor some basic indicators and activity data, and an armband device \cite{lopez2015wearable} utilises optical sensors analysing the concentration of oxygen to obtain respiration rate. To monitor sleep quality, a method \cite{sadek2018nonintrusive} is presented that uses a mat embedded with an optical fiber. The system collects some data during sleep, such as sleeping duration, heart rate, respiration rate and sleeping interruption. Ren et al. \cite{ren2015fine} propose to monitor sleep status using smartphone by placing earphone to record human breathing sound. A wearable neck-cuff system is developed in \cite{rofouei2011non} using microphone, oximetry sensor and accelerometer to monitor sleep. The recent smart bracelets and smart watches \cite{lee2016analysis} have the function of monitoring the quality of sleep. For example, using accelerometers or gyroscopes can obtain real-time motion data and recognise human activity. This method can monitor sleep in real time, and is widely used, although it is uncomfortable due to long-term wear, or sometimes the sensor cannot record due to unconscious movements during sleep.

\subsection{Non-Contact-Based Methods} 
The vision-based approaches \cite{li2016noncontact, scully2011physiological,boccanfuso2012remote} utilise image processing to analyse changes in the human chest and recognise human movement during sleep, which belongs to the  non-contact-based methods. For example, a thermal camera can detect the changes of temperature in nasal \cite{ pereira2015remote} or airflow \cite{murthy2006noncontact} to estimate breathing rate. In addition, in order to reduce complexity, an OSA monitoring system based on infrared cameras \cite{li2016noncontact} is proposed to detect heart rate and respiration rate. For this method, the awareness of privacy protection has become another concern, and the light environment factors have great influence on the performance despite of the contactless method.

The radar-based method, such as Doppler radar \cite{li2009accurate, salmi2011propagation,nguyen2016continuous,park2006single}, ultra wide-band (UWB) radar \cite{salmi2011propagation,cianca2009fm} and frequency-modulated continuous wave (FMCW) radar \cite{adib2015smart,GI2018xiaopeng,2016xiaopeng,ziqian2018}, can penetrate through clothing and quilts to monitor vital signs. For example, a novel model \cite{zhao2017learning} is presented to predict sleep stages, which uses a combination of convolutional and recurrent neural networks to extract features from radio frequency signals and captures their dynamics.  Respiration rate and heart rate can be calculated through the principal component analysis technique based on UWB reflected signals \cite{pittella2017cardiorespiratory}. Moreover,  in  \cite{cho2017novel},  an UWB impulse radar is applied to medical health for detecting heart rate. The authors use ECG data as the ground truth and the overall estimation error is 0.22 $\%$ for heart rate. The Vital Radio system \cite{adib2015smart} uses FMCW radar to measure respiration, which requires unique designed hardware. As a result, commercial respiration monitor radar devices are costly.

The techniques of WiFi communications have great influence on people's lives. The improvement of WiFi device performance has enabled its widespread applications, such as localisation indoors \cite{wu2012fila, wang2014eyes, zhou2015wifi,yang2013rssi,ruichen2018},  posture recognition \cite{wang2015understanding},  human daily activity detection \cite{ohara2017detecting}, fall detection \cite{wang2016wifall} and  sleep quality monitoring. Compared with other wireless communication technologies, WiFi has the advantage in indoor communications due to its transmission quality, adaptability and wide coverage, which is helpful for collecting high quality CSI in complex environments.

The human body can affect signal propagation in physical space, such as reflections or diffraction, as part of physical channel transmission. Respiration, for example, can create a displacement of 4 to 12 millimetres in the chest. It can detect respiration rate through the movement of the chest for WiFi to perceive the human vital signal. WiFi perception technologies can be classified mostly based on received signal strength (RSS) \cite{DiZhai2015} or CSI.

The RSS is coarse-grained channel information, which can be used in indoor localisation, tracking subject and monitoring respiration and heart rate. For example, a method based on RSS \cite{patwari2013breathfinding} is proposed to estimate respiration rate under movement interference and detect its position. The UbiBreathe \cite{abdelnasser2015ubibreathe} system detects respiration rate and apnea through WiFi RSS, requiring a line-of-sight path between the WiFi transmitter and receiver. The BreathTaking \cite{patwari2013monitoring} system can estimate breathing rate with WiFi RSS, which can be improved by using directional antennae. However, the RSS method is affected by two elements: occlusion, electromagnetic environment variations, etc. 

The RSS provides coarse-grained channel information that can be used for indoor localization, subject tracking, and respiration and heart rate monitoring. For example, in \cite{patwari2013breathfinding} a method based on RSS is presented to estimate respiration rate and detect position under movement interference. The UbiBreathe system \cite{abdelnasser2015ubibreathe} uses WiFi RSS to detect breathing rate and apnea, however it requires a direct line of sight between the WiFI transmitter and receiver. With WiFi RSS, the BreathTaking system \cite{patwari2013monitoring}  can measure breathing rate, which can be improved with directional antennae. However, the detection accuracy of the RSS method is largely affected by occlusion, electromagnetic environment changes, and so on.

Recent studies demonstrate a huge improvement in WiFi perception by using fine-grained information instead of coarse-grained channel information to track human vital signals. CSI is fine-grained channel information which is more suitable for recording small scale activity. 

Liu et al. \cite{liu2018monitoring}, for example, use the amplitude of CSI to determine the respiration rate and heart rate for one or two people. They calculate heart rate and respiration rate using the power spectral density, and they compare system performance with different distances between the transmitter and receiver. The BreathTrack system \cite{zhang2019breathtrack} uses the phase information of CSI to measure the human breathing signal. The CardioFi \cite{khamis2018cardiofi} system can obtain heart rate by calculating the difference in phase information between two antennae and describe a novel algorithm for selecting sub-carriers. 

The Mo-Sleep \cite{li2016mo} device monitors sleep in two ways. The motion detection segment uses the amplitude and phase information from CSI to identify movement during sleep, while the breathing monitoring section uses principal component analysis to calculate the breathing rate. The WiFi-Sleep \cite{yu2021wifi} system uses a deep learning algorithm to categorize four different types of sleep stages. FullBreathe\cite{zeng2018fullbreathe}  detects human respiration using the complementary of amplitude and phase information of CSI. The authors of \cite{lee2018design} use WiFi devices to recognise breathing and heart rate using the Dynamic Time Warping technique.  Monitoring the breathing rate in the car can also be achieved using CSI \cite{hussain2021vehicle}. Amplitude and phase information is used to predict respiratory rate in a stationary vehicle, and it is discovered that multiple antennae can benefit system performance.

A non-contact audio method is proposed in \cite{deng2017decision}. It extracts the snore and breath features from audio signals to estimate the four sleep stages by using a decision tree. \cite{dafna2016estimation} presents a method using a whole night audio information to detect REM, NREM and awake stages. \cite{dafna2015sleep} uses Adaboost to classify sleep and wakefulness by acoustic features extracted from breathing sounds. In addition, to monitor sleep apneas, a smartphone based on active sonar \cite{nandakumar2015contactless} has been developed, to detect the movement of the chest while breathing through the reflected signal.

Besides, the Fresnel Zone model \cite{gu2019wifi,wang2016human, wu2017device} can also be used to monitor breathing rate and heart rate. \cite{wang2016human} found that CSI can perform better in the middle of the Fresnel zone than at the boundaries.


\section{System Overview}
\begin{figure}
\centerline{\includegraphics[width=0.45\textwidth]{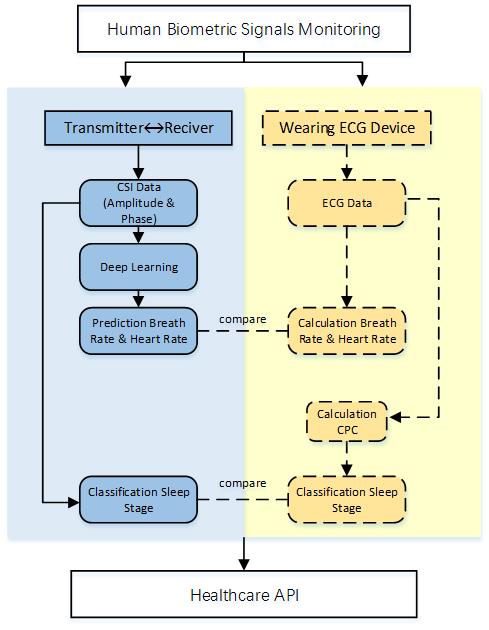}}
\caption{The overview of the system}
\label{fig}
\end{figure}

In this paper, we propose to monitor respiration and heart rate by capturing WiFi signals so as to classify sleep stages. The calibration is carried out  using a wearable ECG device. As illustrated in Fig.~\ref{fig}, the system can be divided into two parts: the first part is to detect human biometric signals based on WiFi devices and the other uses an ECG device to record the ground truth data for respiration rate and heart rate estimation.

The CSI data is collected by a 802.11n WiFi device (the TP-Link TL-WDR4300 wireless router). The flowchart for WiFi based data processing is illustrated by the blue part of Fig.~\ref{fig}. Based on the CSI time series data, we can extract the amplitude and phase information of the WiFi signal. The CSI data is then sent into our developed Heart Rate and Respiration Rate Network (H3RN) for heart rate and respiration rate estimation. 
Meanwhile, for the yellow section in Fig.~\ref{fig} the participants need to wear our developed wearable single-lead ECG device \cite{web,xucun2020WCNC,menglu2021,zijiao2021,wang2022wearable} to record the ECG signal. Based on the ECG data, we then calculate the instant heart rate and the ECG derived respiratory (EDR) as our ground truth data. The detailed description for heart rate and respiration rate estimation is given in Section \uppercase\expandafter{\romannumeral4}.

For sleep stage classification, we developed two neural networks (the W2SN and CPC Neural Network) to classify four stages: wake, REM sleep, light sleep, and deep sleep. The input for the W2SN is the raw CSI, and the input for the CPC network is the CPC signal. Section \uppercase\expandafter{\romannumeral5} provide detailed description about these networks. Finally, we compare our results to those derived using ground truth data to assess the whole system's performance.


\section{Heart Rate and Respiration Rate Estimation}

For the heart rate and respiration rate estimation, we first present data pre-processing, containing CSI data and ECG data. After that, we introduce a single input multiple output CNN to estimate heart rate and respiration rate, which utilise the inner relationship between these two rates.

\subsection{Data Processing}
\paragraph{CSI Data Processing}
It is necessary to point out that each value of CSI is a complex number while in most of the cases, the neural network takes real values as inputs. As such, we can break down the CSI into amplitude and phase data. Mathematically, the $n$th value $G_n$ of the CSI can be expressed as:

\begin{equation}\label{Eqn:Amp}
    G_n=A_n\mathrm{e}^{jP_n}
\end{equation}
where $A_n$ and $P_n$ are the amplitude and the phase of $G_n$, respectively. 

\begin{figure}
\centerline{\includegraphics[width=0.53\textwidth]{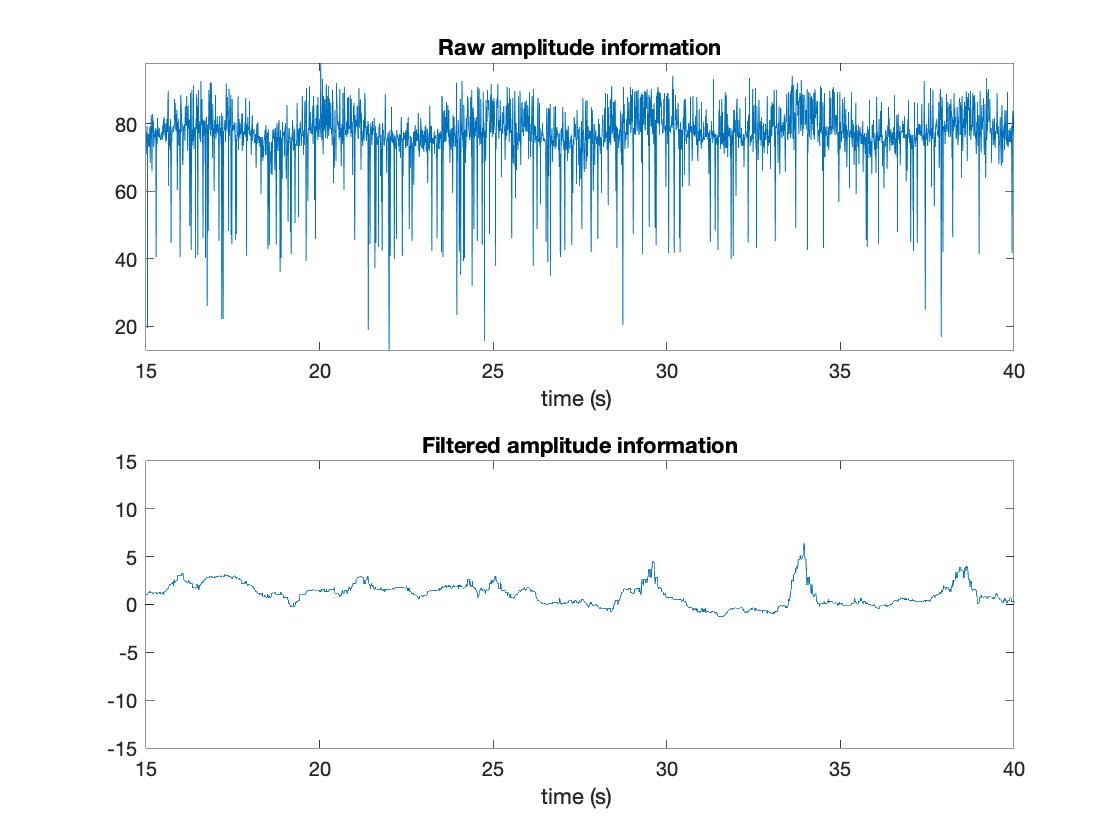}}
\caption{Filtered amplitude information}
\label{Fig:amplitude}
\end{figure}

\begin{figure}
\centerline{\includegraphics[width=0.53\textwidth]{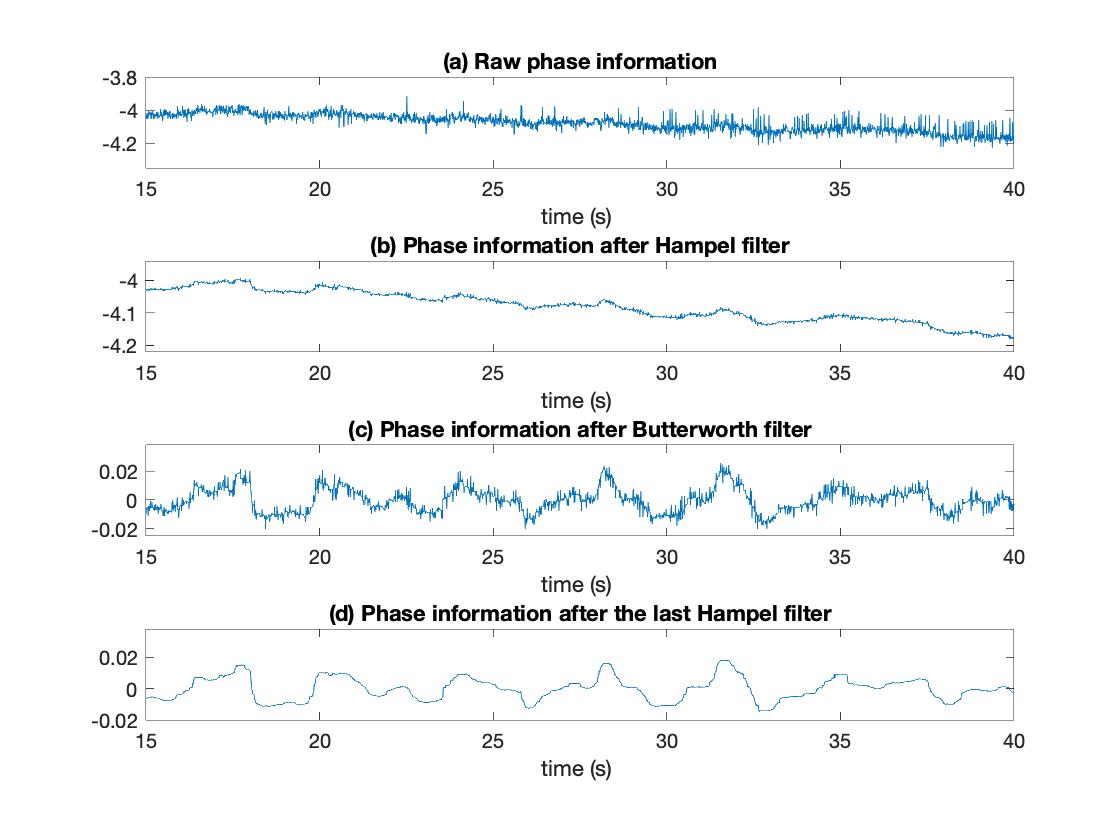}}
\caption{Filtered phase information, (a) Raw phase information, (b) Phase information after Hampel filter, (c) Phase information after Butterworth filter, (d) Phase information after the last Hampel filter}
\label{Fig:phase}
\end{figure}

The amplitude and phase information of the CSI for one sub-carrier are shown in Figs.~\ref{Fig:amplitude} and \ref{Fig:phase}, respectively. We can extract the amplitude information from one of the receiving antennas. In this paper, we extract the amplitude information from the first receiving antenna and process data using the Butterworth and Hampel filters to remove outliers and noise, as shown in Fig.~\ref{Fig:amplitude}. To cancel the phase noise, we first calculate the phase difference between the receiving antennas. In this way, we can offset a random phase shift. The raw phase information is shown in Fig.~\ref{Fig:phase}a. In order to obtain smooth curve of phase information, we need to filter out the outliers and remove noise. In this work, we use the Hampel filter  to remove the outliers. The window size  and the threshold coefficient of the filter are set to 51 and 0.4, respectively. Fig.~\ref{Fig:phase}b shows the filtered phase. The Butterworth filter is capable of removing both direct current (DC) component and very low frequency noise. The cut-off frequency is set to 0.2. The result is illustrated in Fig.~\ref{Fig:phase}c. Finally, we apply a secondary Hampel filter to further remove noise. The filtered phase information is shown in Fig.~\ref{Fig:phase}d.

After that, we concatenate the amplitude and phase information to obtain our CSI matrix $\boldsymbol{Q}$, which is a three dimensional matrix that serves as the neural network's input. The dimension of $\boldsymbol{Q}$ is a $\left(t \times  f_s\right)$-by-$m$-by-2 where $t$ is the sampling duration in seconds, $f_s$ is the sampling rate and $m$ represents the number of sub-carriers which is set to 56 in this system, 2 means using both the amplitude and phase information.
    
\paragraph{ECG Data Processing}

\begin{figure}
\centerline{\includegraphics[width=0.50\textwidth]{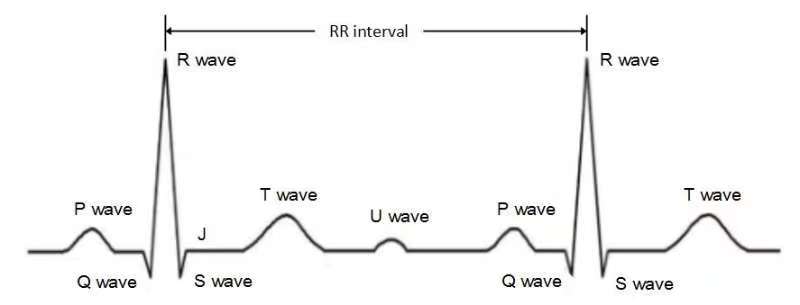}}
\caption{RR interval}
\label{Fig:RR}
\end{figure}

The ECG data is collected using our developed wearable ECG device, called irealcare (version 2) \cite{web}. Based on the collected ECG data, we can calculate the instant heart rate and the EDR. To this end, we need to extract the RR interval, which is the interval between two adjacent R waves based on the value of R peak and their corresponding time series. Fig.~\ref{Fig:RR} shows the RR interval and the QRS complex of ECG signals.

\begin{figure}
\centerline{\includegraphics[width=0.53\textwidth]{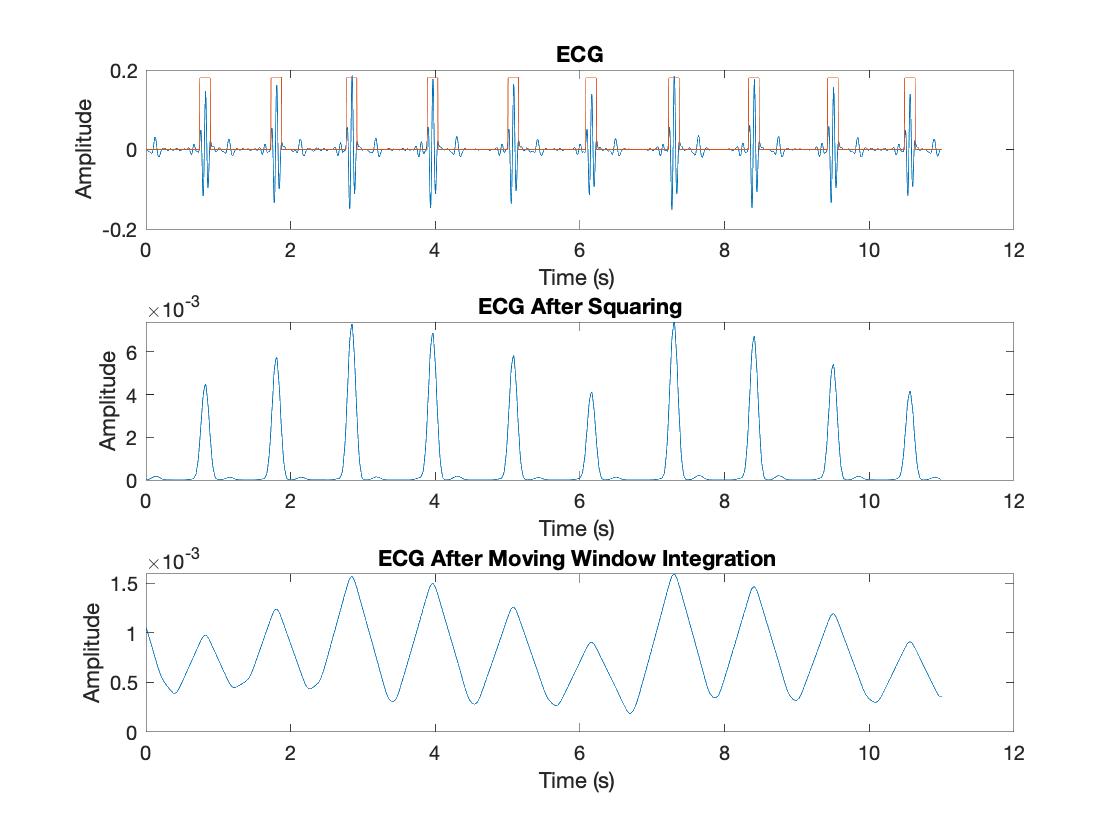}}
\caption{The filtering process of sleep ECG signal}
\label{Fig:The Process of Filtering ECG Signal}
\end{figure}

We first obtain the QRS complex from the filtered ECG signal to extract the RR interval, as $x\left[n\right]$. Then, we normalise the ECG data to remove the influence of signal gain, as ${{x_{normalised}}\left[n\right]}$. After normalization, a band-pass filter is used to remove baseline drift, as $q\left[n\right] $=$ BPF({{x_{normalised}}\left[n\right]})$. Then, the signal is amplified, as $p\left[n\right]$=$q\left[n\right]^{2}$, while the values of the Q wave and S wave are easily extracted. Two windows are used to respectively remove the noise and detect the QRS complex. For effective noise removal, the window size is set to 97 milliseconds (24 samples at 250 Hz). For detecting the QRS complex, the window size is set to 611 ms (153 samples for 250 Hz) \cite{elgendi2013fast}. By comparing with the threshold, the QRS complex can be detected as shown by the red line in Fig.~\ref{Fig:The Process of Filtering ECG Signal}. QRS complex can be determined in the selected range. Besides, the filtering process of sleep ECG signal is shown in Fig.~\ref{Fig:The Process of Filtering ECG Signal}.  


After extraction of the QRS complex, heart rate can be obtained by counting the number of R waves per unit time. It can be calculated by using the RR interval, as shown in Fig.~\ref{Fig:RR}. The formula of heart rate can be expressed as:

\begin{equation}\label{Eqn:snr}
    HeartRate=\frac{60} {RR\,Interval}
\end{equation}

According to \cite{lipsitz1995heart}, EDR can be obtained by calculating the area under each QRS complex which is consistent with the change trend of the respiratory cycle. The respiration signal derived from the ECG can avoid the burden on the human body caused by wearing additional hardware equipment. When a person inhales or exhales, the air in their lungs is filled or emptied, and this is how EDR is acquired. This phenomenon leads to impedance distributed on the chest surface or the chest volume varying. These changes can deflect the axis of ECG. Therefore, measuring the angle of the vector of ECG axis can obtain respiration signal. 

\begin{figure}
\centerline{\includegraphics[width=0.53\textwidth]{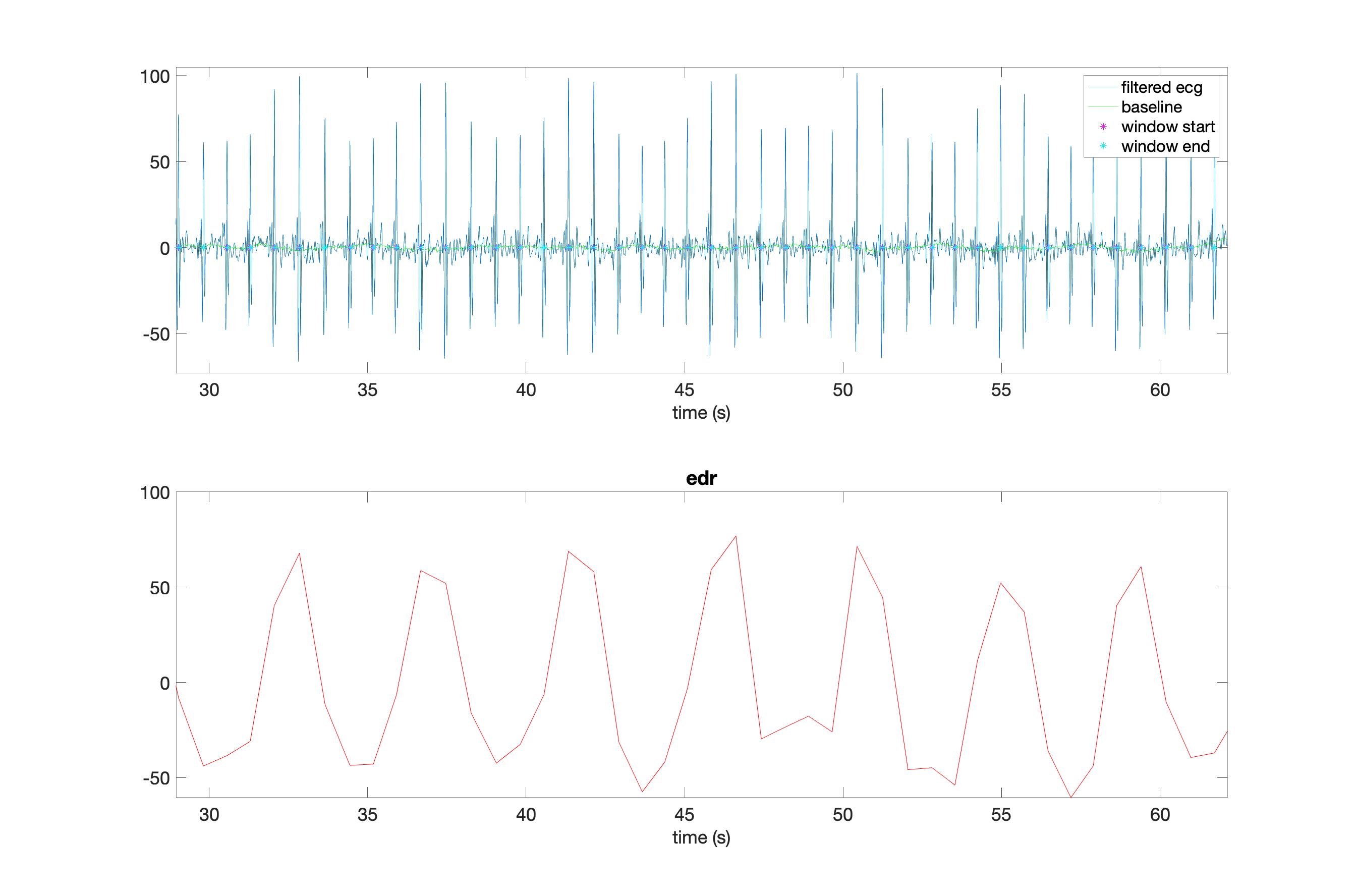}}
\caption{ECG signal and EDR signal}
\label{Fig:ECG Signal and EDR Signal}
\end{figure}

There are several techniques which can be utilized to obtain the respiration signal from ECG. The Area method, by calculating of the QRS complex area, can measure the angle of the vector of the ECG axis. First, the baseline drift for the ECG signal should be removed. Second, the area of the QRS complex can be calculated by the fixed moving window, which is set to the distance between P, Q and J points for QRS complex. The QRS complex area is proportional to the amplitude of ECG signal. The ECG signal and the corresponding EDR signal are shown in Fig.~\ref{Fig:ECG Signal and EDR Signal}.


\subsection{Heart Rate and Respiration Rate Network (H3RN)}
 Using the intrinsic link between heart rate and respiration rate, the H3RN is a single input two output CNN and can measure both rates at the same time. This is inspired by the fact that heart rate and respiration rate are always related to each other \cite{wallin2010relationship} and, therefore, they can be jointly estimated. 

\begin{figure}
\centerline{\includegraphics[width=0.56\textwidth]{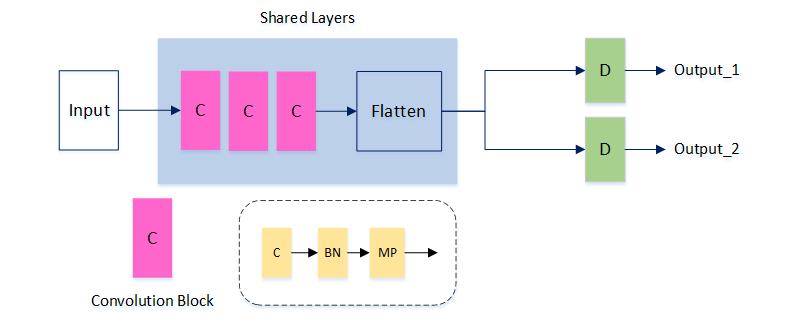}}
\caption{H3RN structure}
\label{Fig:Neural_network_structure}
\end{figure}

The structure of the H3RN is depicted in Fig.~\ref{Fig:Neural_network_structure}. It consists of three layers: share, flatten and fully connect (FC). The share layers in the proposed neural network contain three convolution blocks.  This network aims to use more information from CSI (both amplitude and phase) to improve accuracy and reduce computational complexity in the training network. 

In the share layers, the first convolutional block includes a 5-by-5 convolutional layer, a batch normalization (BN) layer and a 4-by-4 max pooling (MP) layer. The second convolutional block contains a 3-by-3 convolutional kernel, a BN layer, and a 2-by-2 pooling size. The last convolutional block has a convolutional layer with a 2-by-2 kernel and a 3-by-3 pooling size. The function of the flatten layer is to flatten the CNN output into one dimension. The purpose of building shared layers is to reduce the computational complexity of the whole network. After the FC layer, the respiration rate and heart rate can be estimated.    

\section{Sleep Stage Monitoring}
\begin{figure}
\centerline{\includegraphics[width=0.48\textwidth]{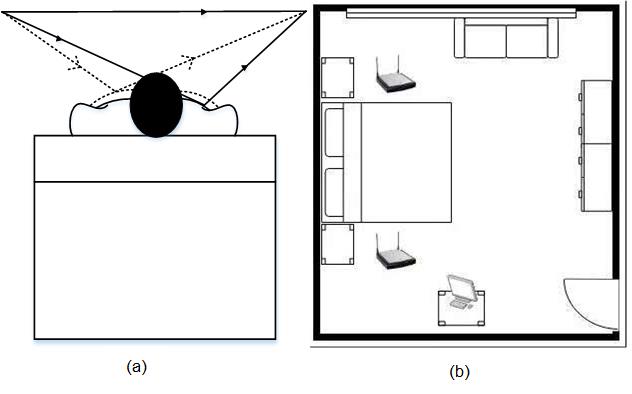}}
\caption{WiFi based sleep stages detection deployment. (a) Front view of WiFi sleep monitoring. (b) Sleep environment setup.}
\label{Fig:Front view of WiFi sleep monitoring}
\end{figure}

Sleep stages are usually classified into three categories: wake, NREM and REM. In \cite{berry2012aasm, roth2009slow}, the $N_1$ and $N_2$ stages from NREM are often considered as light sleep and the $N_3$ stage usually represents deep sleep from NREM. Sleep may now be divided into four stages: wake, light sleep, deep sleep and REM. Fig.~\ref{Fig:Front view of WiFi sleep monitoring} shows the deployment for sleep stage detection based on WiFi. 

In this section, we develop two classification approaches: the WiFi approach and the CPC approach. For the first one, we develop the W2SN network and the input data is the CSI matrix without further processing. For the second one, we design the CPC network with the CPC signal as the input data.

The W2SN can simplify existing sleep stage classification approaches such as those described in \cite{li2016mo,yu2021wifi}. This is  because the CSI is influenced not only by large-scale movement (body movement) but also by tiny-scale movement (chest movement). Human body-movement data, respiration data and heart rate during sleep are already included in the CSI data. This also explains why we do not use the motion-sensing module in our system. We directly use the CSI matrix without further processing as input for classification. 


To obtain the CPC signal for the CPC neural network, we first normalize the ECG data and remove the baseline drift by a bandpass filter. By squaring the signal, the R wave can be amplified and noise suppressed. Then, the range of QRS complex can be located with two moving windows. The QRS complex in the selected range can be determined by comparing it with the threshold. We can convert the QRS complex into a time series of EDR signal and synchronous heart rate variability (HRV) using the QRS complex extraction. The RR interval is used to calculate HRV. The method described in \cite{lipsitz1995heart} is used to retrieve the relevant EDR signal. CPC signal can be calculated by the coherence of the HRV and EDR signals, which can be described by the cross-spectrogram. In addition, the referenced labels for four sleep stages are recorded through a smart wristband Fitbit (inspire HR). 

\subsection{W2SN}
\begin{figure}
\centerline{\includegraphics[width=0.53\textwidth]{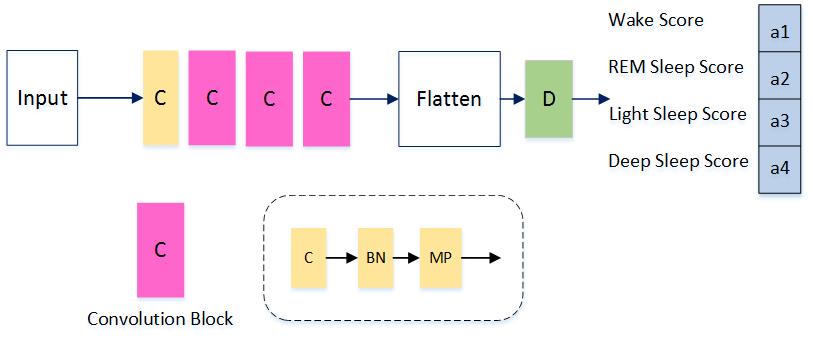}}
\caption{The structure of the W2SN. Each convolutional block consists of a convolutional layer, a BN layer and a MP layer.}
\label{Fig:The structure of Sleep Network}
\end{figure}

The input for the neural network based on WiFi is the CSI matrix $\boldsymbol{Q}$. Fig.~\ref{Fig:The structure of Sleep Network} describes the structure of the W2SN. The first layer is a 5-by-5 convolutional layer, followed by three consecutive convolution blocks. The first convolution block includes a convolutional layer with a 3-by-3 convolutional kernel, a 40-by-16 pooling layer and BN which can prevent overfitting. The second convolution block contains a 2-by-2 convolutional kernel, a 20-by-4 pooling layer and BN. In the 
third convolution block, the pooling size is 12-by-2. After passing the softmax classifier, we can obtain the score of probability with the four types of sleep stages and select the category with the highest probability. 

According to \cite{lin2016sleepsense, yu2021wifi,wilde1983rate}, the sleep stages are related to the respiration and body movements. For instance, \cite{wilde1983rate} claims that sleep can be measured through body movements due to the strong correlation between sleep stage and body movement rate. As mentioned early, we do not use the motion-sensing module in our system. Rather, we directly use the CSI matrix without further processing as input for classification. This is because body-movement data, respiration data and heart rate during sleep are already included in the CSI data.

\subsection{CPC Neural Network}
For the neural network based on CPC algorithm, the network consists of four convolution blocks. Each convolution block comprises a convolution layer, a BN and a pooling layer. The convolutional kernel sizes for the first, the second, and the third convolutional blocks are set to 5, 3 and 2, respectively, and the pooling size is set to 2.

\section{Performance Evaluation}

We evaluate system performance from two aspects: (1) the accuracy of heart rate and respiration rate estimation, and (2) the accuracy of sleep stage classification.
\subsection{Hardware Setup}
\begin{figure}
\centerline{\includegraphics[width=0.48\textwidth]{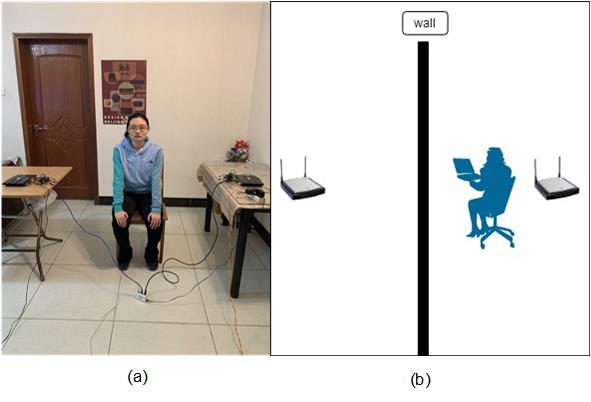}}
\caption{Experiment environment setup. (a) Direct line-of-sight (LOS) environment. (b) Non-line-of-sight (NLOS, through the wall) environment.}
\label{Fig:Experiment Environment Setup}
\end{figure}

The WiFi based system has two identical 802.11n WiFi devices (TP-Link TL-WDR 4300 wireless routers) acting as a transmitter and a receiver respectively. Both use 20 MHz bandwidth and operate at 2.4 GHz. The firmware is modified on OpenWRT in order to extract CSI with 56 sub-carriers from received signals. For data processing, we use a laptop (Lenovo L460). Fig.~\ref{Fig:Experiment Environment Setup} shows the environment setting for LOS and NLOS (through the wall) scenarios in the case of heart rate and respiration rate estimation. The receiver and transmitter are placed at two sides of a participant and the distance between them is two metres. For sleep stage monitoring, the scenario is depicted in Fig.~\ref{Fig:Front view of WiFi sleep monitoring}. In this paper, ground truth data for monitoring respiration rate and heart rate are detected by a wearable ECG device irealcare (version 2). In addition, we use smart wristband Fitbit (inspire HR) to record the label for the sleep stage.

\subsection{Heart Rate and Respiration Rate Estimation}
\begin{figure}
\centerline{\includegraphics[width=0.5\textwidth]{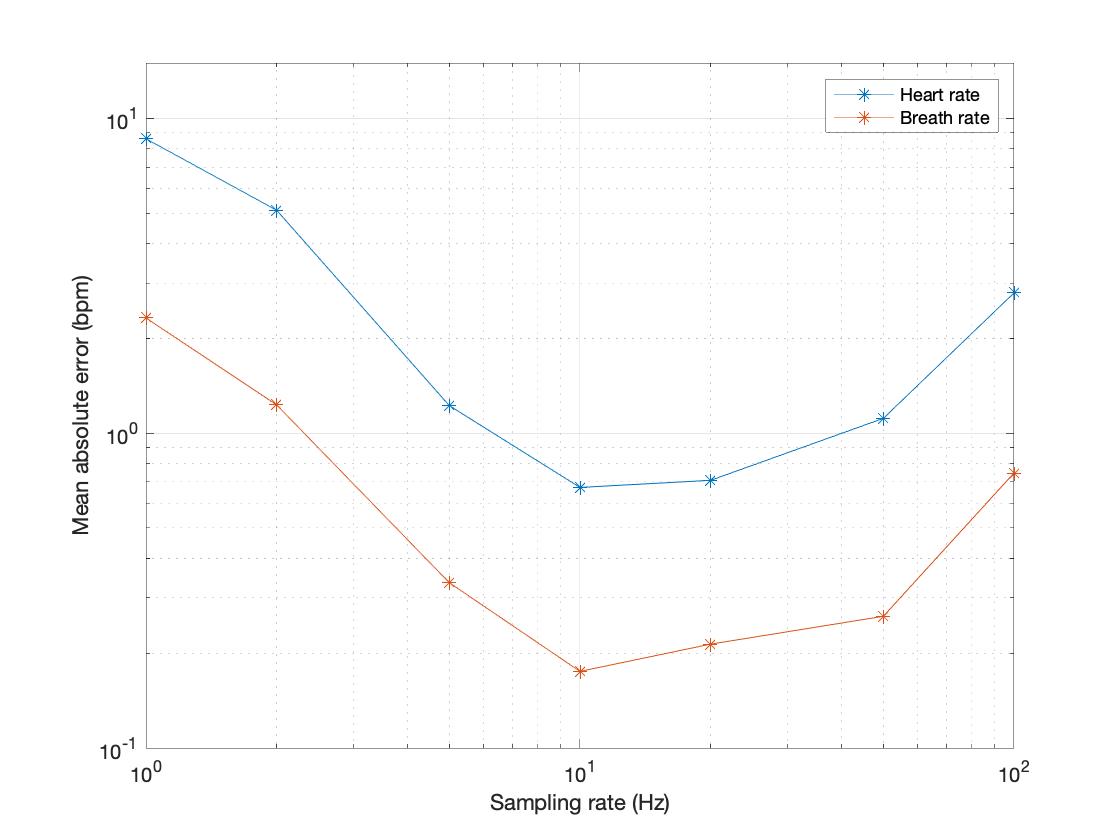}}
\caption{Performance under different sampling rates}
\label{Fig:The Relationship between Sampling rate and MAE}
\end{figure}

\paragraph{Effect of Sampling Rate and Sampling Duration} We resample the CSI data and set them to 1 Hz, 2 Hz, 5 Hz, 10 Hz, 20 Hz and 50 Hz when the sampling duration is 50 s. Fig.~\ref{Fig:The Relationship between Sampling rate and MAE} shows the relationship between the sampling rate and the mean absolute error (MAE) of the heart and respiration rate. We utilize MAE as an indicator to evaluate the performance of heart rate and respiration rate estimation. 

As shown in Fig.~\ref{Fig:The Relationship between Sampling rate and MAE}, the MAEs for both heart rate and respiration rate estimation generally show a downward trend at first and then tend to increase as the sampling rate increases. At the sampling rate of 10 Hz, the MAE of heart rate is 0.6042 beats per minute (bpm). Beyond this point, the trend of heart rate MAE increases. For respiration, the MAE of respiration rate is 0.2 bpm at the 10 Hz sampling rate, which is a turning point for the whole range of sampling rates. When the sampling rate is set to 10 Hz, the system exhibits high accuracy in estimating respiration rate and heart rate. 

This is due to the fact that the desired signals focus on low frequency range (i.e. below 2 Hz). The range of respiration rate is usually $10\textendash37$ bpm. It corresponds to the respiration signal frequency under 1 Hz. The adults' respiration rate is usually at the range of $10\textendash14$ bpm and babies' respiration rate  is usually around 37 bpm \cite{liu2018monitoring}. For people at rest, the heart rate is $60\textendash80$ bpm \cite{liu2018monitoring}. This corresponds to under $1\textendash1.33$ Hz. When reducing the sampling frequency, the cut-off frequency of anti-aliasing filter will be reduced accordingly, which is equivalent to applying a low-pass filter with narrower pass band. The anti-aliasing filter may ensure that the signal bandwidth satisfies the Nyquist sampling theorem, preventing frequency spectrum aliasing.

In this case, the noise power is reduced (due to smaller signal bandwidth), while the power of desired signal keeps almost the same and therefore the  signal-to-noise ratio (SNR) increases. The received SNR can be expressed as:

\begin{equation}\label{Eqn:snr}
	SNR=\frac{S}{N_0B}
\end{equation}
where $S$ is the power of signal, $B$ is the bandwidth and $N_0$ represents the noise power density. This formula clearly shows that as the bandwidth increases, SNR decreases, adversely affecting the system performance. 

\begin{figure}
\centerline{\includegraphics[width=0.53\textwidth]{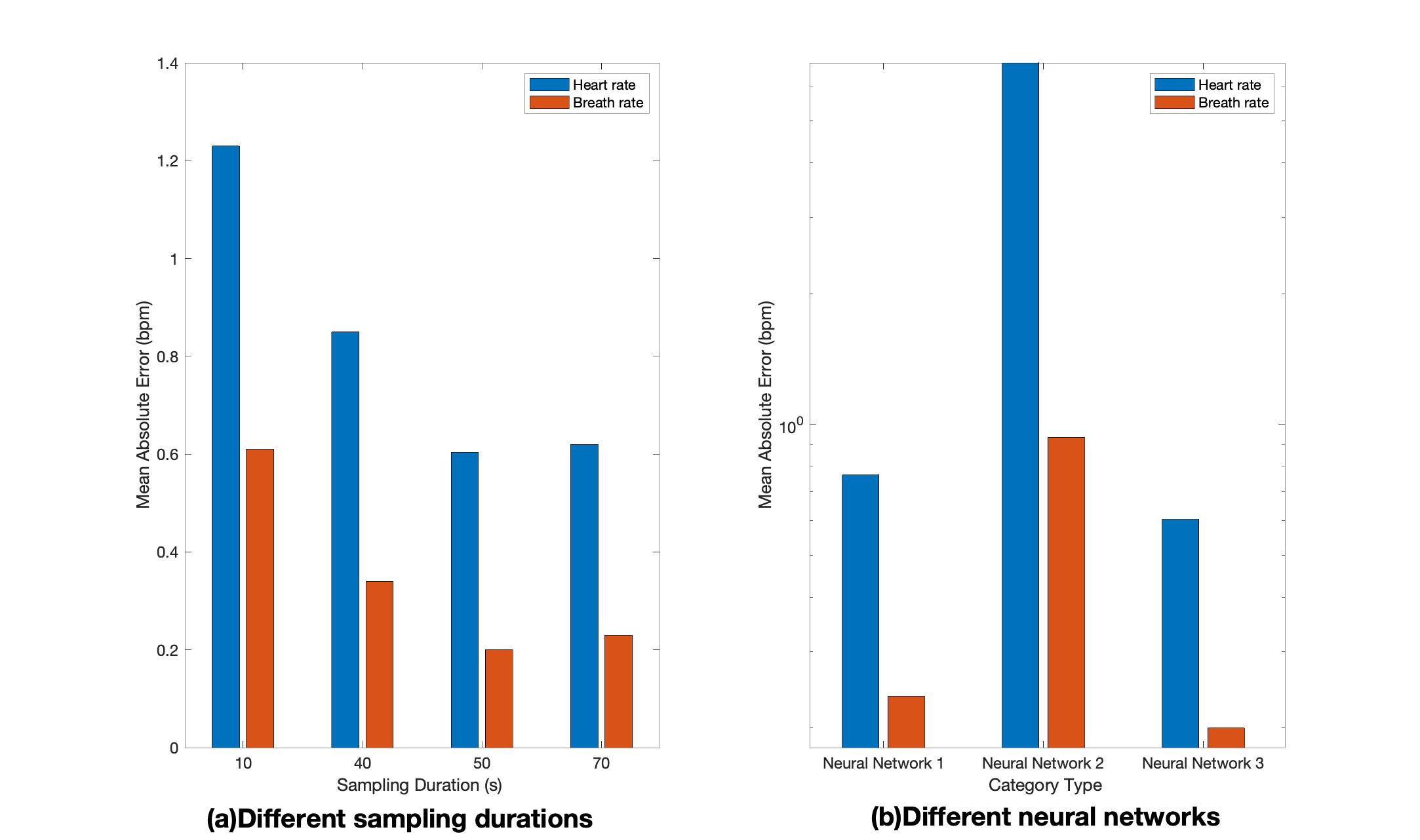}}
\caption{Performance under different parameters. (a)Performance under different sampling duration. (b)Performance under different neural networks.}
\label{Fig:3}
\end{figure}

Upon selection of the best sampling rate for this system, we set different sampling duration of 10 s, 40 s, 50 s and 70 s. The sample data is a three dimension matrix of size $t$ by $m$ by 2. $t$ represents the sample duration. $m$ is the number of sub-carriers, which is set to 56. As shown in Fig.~\ref{Fig:3} (a), as the sampling duration becomes longer, the CSI data contains more information, which can effectively improve the accuracy. When the sampling duration is between 10 s and 50 s, the trend of MAE in this system decreases. However, as the sampling duration continues to increase, the MAE remains stable. It is because the background noise affects the whole system performance after 50 s.


\begin{table}\caption{Comparison of different neural network structures }\label{table:setup}
\centering
{\centering}
\begin{tabular}{|c|c|c|c|c|c|c|}
\hline
Types & Error for respiration & Error for heart rate\\
\hline\hline

{Chen et al. \cite{chen2018deepphys}} & 3.02 bpm & N/A  \\  
\hline
{PhysNet \cite{yu2019remote}} & N/A &  2.57 bpm   \\  
\hline
{Meta-rPPG \cite{lee2020meta}} & N/A &  3.62 bpm   \\  

\hline
{Neural network 1} & 0.2368 bpm & 0.7647 bpm  \\ 
\hline
{Neural network 2} & 0.9337 bpm &  7.545 bpm   \\  
\hline
{Neural network 3} & 0.2 bpm &  0.6042 bpm  \\ 
i.e. H3RN &&\\

\hline
\end{tabular}
\end{table}

We also compared with various neural network structures. The results are shown in Table~\ref{table:setup}, in which Neural network $1$ contains one convolutional layer, neural network $2$ contains five convolutional layers and neural network $3$ is our developed neural network structure based on WiFi,  i.e., the H3RN network depicted in Fig. \ref{Fig:Neural_network_structure}, which contains three convolutional layers. The estimation error of neural network 1 for respiration rate is 0.2368 bpm, and for heart rate is 0.7647 bpm. The estimation error of neural network 2 for respiration rate is 0.9337 bpm and for heart rate is 7.545 bpm. We compare these three type neural networks in Fig.~\ref{Fig:3} (b). Overall, the figure shows that neural network 3, i.e., the H3RN network, performs better than the other two networks. In Table~\ref{table:setup}, we also listed the estimation error of the heart rate for PhysNet \cite{yu2019remote} and Meta-rPPG \cite{lee2020meta} and of the respiration rate  obtained from \cite{ni2021review}. It can be seen that our proposed H3RN network gives the best performance. In the next section, we explicitly introduce the performance for neural network 3, i.e. our proposed network structure based on WiFi.

\begin{figure}
\centerline{\includegraphics[width=0.53\textwidth]{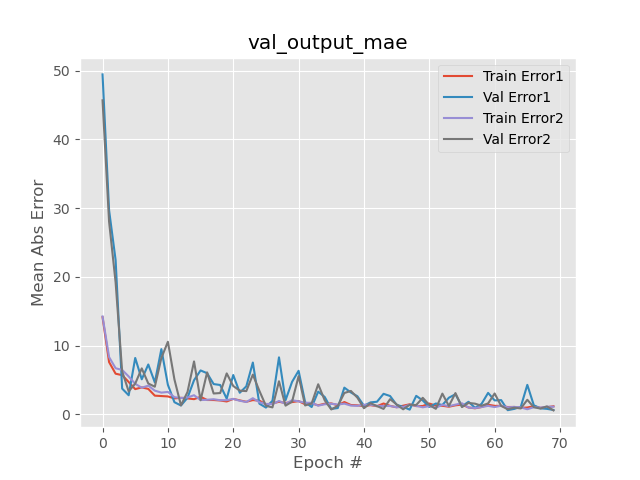}}
\caption{Training performance}
\label{Fig:Training Performance}
\end{figure}

\begin{figure}
\centerline{\includegraphics[width=0.53\textwidth]{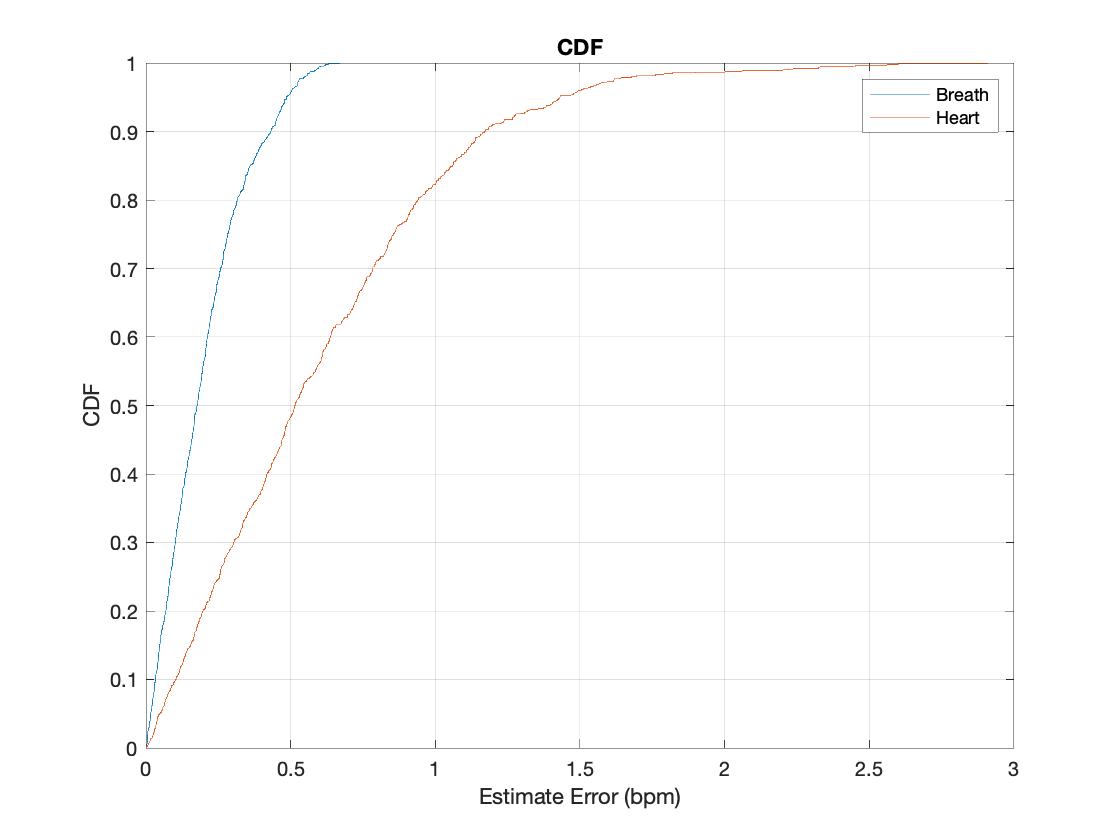}}
\caption{CDF of estimation error}
\label{Fig:CDF of Estimation Error}
\end{figure}

\paragraph{Performance of Heart Rate and Respiration Rate Estimation} Fig.~\ref{Fig:Training Performance} describes the MAE of respiration and heart rate with 50 s sampling duration. Val Error1 with blue line shown in the figure represents the validation error of heart rate. Val Error2 with purple line is the validation error of respiration rate. As the epoch increases, the MAE of validation becomes stable. It clearly shows that the value of MAE for low to medium epochs fluctuates. After $40$ epochs, the system error remains almost constant. For both the heart rate and respiration rate error, we calculate the cumulative distribution function (CDF) shown in Fig.~\ref{Fig:CDF of Estimation Error}.

Finally, we discovered that the MAE values for respiration rate and heart rate estimation are 0.2 bpm and 0.6042 bpm, respectively, as shown in Table~\ref{table:setup}. The accuracy of the system measured by mean absolute percentage error for respiration and heart rate is 99.109$\%$ and 98.581$\%$, respectively. Comparing to existing studies \cite{gu2019wifi,li2016mo}, e.g., the estimation accuracy for respiration of 90.75$\%$ in \cite{li2016mo}  and 96.636$\%$ for respiration   and 94.215$\%$ for heart rate in \cite{gu2019wifi}, these results are encouraging. In addition, we also detect the heart rate and respiration rate in an through wall environment as shown in Fig.~\ref{Fig:Experiment Environment Setup} (b). The through wall (NLOS) environment has one wall in between the receiver and the transmitter. The summary of the experiment results is given in Table ~\ref{table:sum}. In comparison to the LOS scenario, it is clear that there is a performance degradation.

\begin{table}\caption{Summary results for LOS and NLOS }\label{table:sum}
\centering
{\centering}
\begin{tabular}{|c|c|c|c|c|c|c|}
\hline
  & Accuracy for respiration & Accuracy for heart rate\\
\hline\hline
{LOS} & 99.109$\%$ & 98.581$\%$  \\ 
\hline
{NLOS} & 98.2$\%$ & 92.9$\%$    \\  

\hline
\end{tabular}
\end{table}

\begin{figure}
\centerline{\includegraphics[width=0.5\textwidth]{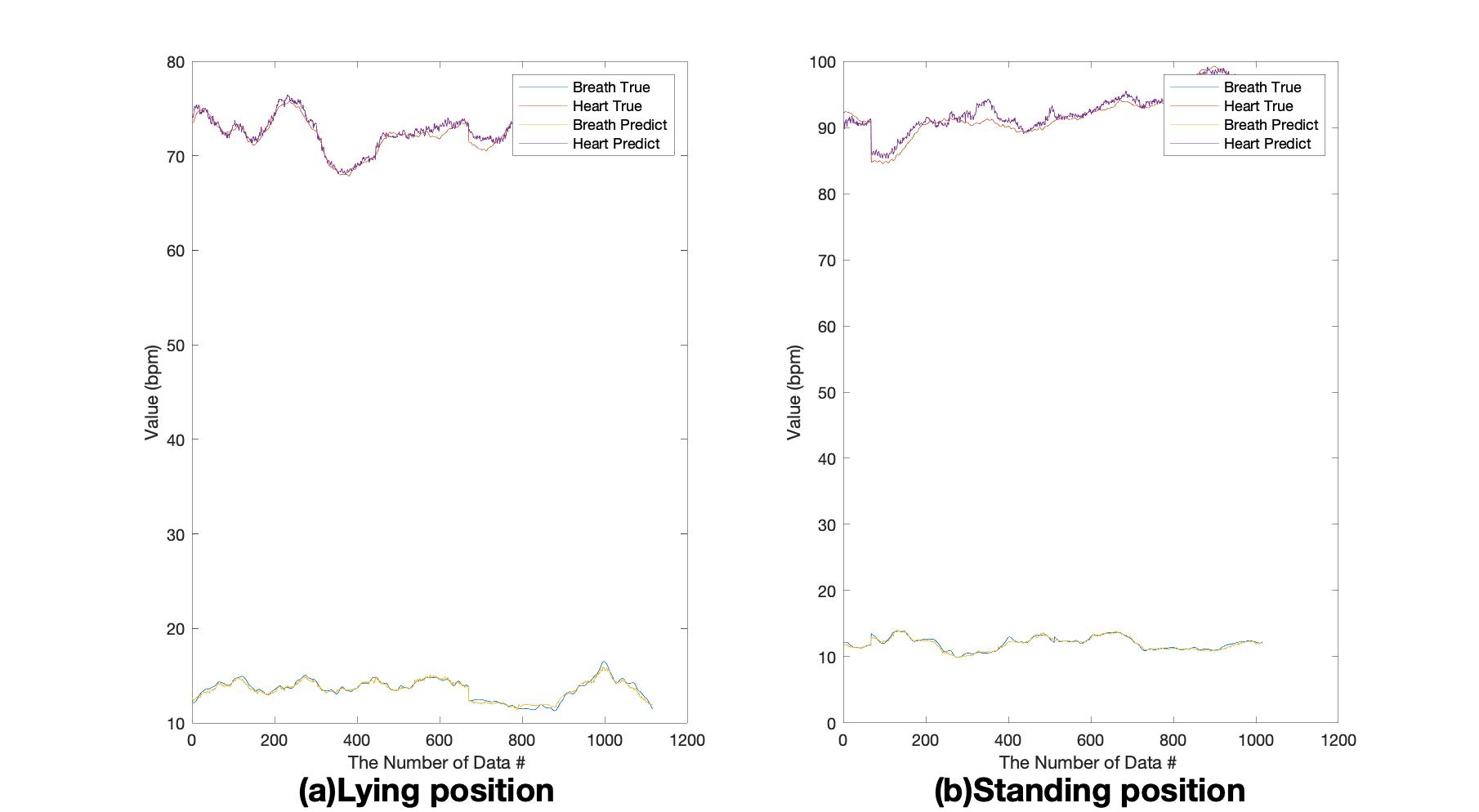}}
\caption{The estimation and ground truth value under different posture. (a) The estimation and ground truth value while lying. (b) The estimation and ground truth value while standing.}
\label{Fig:Predict and Reference Value of Lay}
\end{figure}

\paragraph{Performance of Different Postures} We also conduct an experiment in which we monitor respiration and heart rate while lying in bed and while standing. Fig.~\ref{Fig:Predict and Reference Value of Lay} (a) and Fig.~\ref{Fig:Predict and Reference Value of Lay} (b) show the estimation curve and reference value for respiration rate as well as heart rate under different postures, respectively. Fig.~\ref{Fig:Predict and Reference Value of Lay} (a) shows the respiration rate and heart rate while lying (supine). The heart rate estimation value is shown by the red line, which is almost identical to the green dashed line representing the heart rate obtained from the ECG device. The average estimation value of heart rate is 72.4661 bpm. The average true heart rate is 72.069 bpm. The blue line are the estimation value of respiration rate. The average estimated respiration rate is 13.5293 bpm. The average true respiration rate is 13.5823 bpm.

\begin{table}\caption{Summary of results}\label{table:conclusion1}
\centering
{\centering}
\begin{tabular}{|c|c|c|c|c|c|c|}
\hline
Posture & Accuracy of respiration & Accuracy of heart rate\\
\hline\hline
{Sitting position} & 99.109$\%$ & 98.581$\%$  \\ 
\hline
{Lying position} & 99.272$\%$ & 98.591$\%$    \\  
\hline
{Standing position} & 99.1032$\%$  & 98.666$\%$   \\ 

\hline
\end{tabular}
\end{table}

Fig.~\ref{Fig:Predict and Reference Value of Lay} (b) presents the respiration rate and heart rate while standing. For standing, estimated average heart rate is 92.5696 bpm. The average true heart rate is 92.0795 bpm. The average estimated respiration rate is 11.9362 bpm. The average true respiration rate is 12.0178 bpm. Standing resulted in a higher heart rate and respiratory rate as compared to lying down. People's heart rates and respiration rates are faster when they stand than when they lie down. In reality, it shows that the data we estimate is in line with common sense.

\begin{figure}
\centerline{\includegraphics[width=0.53\textwidth]{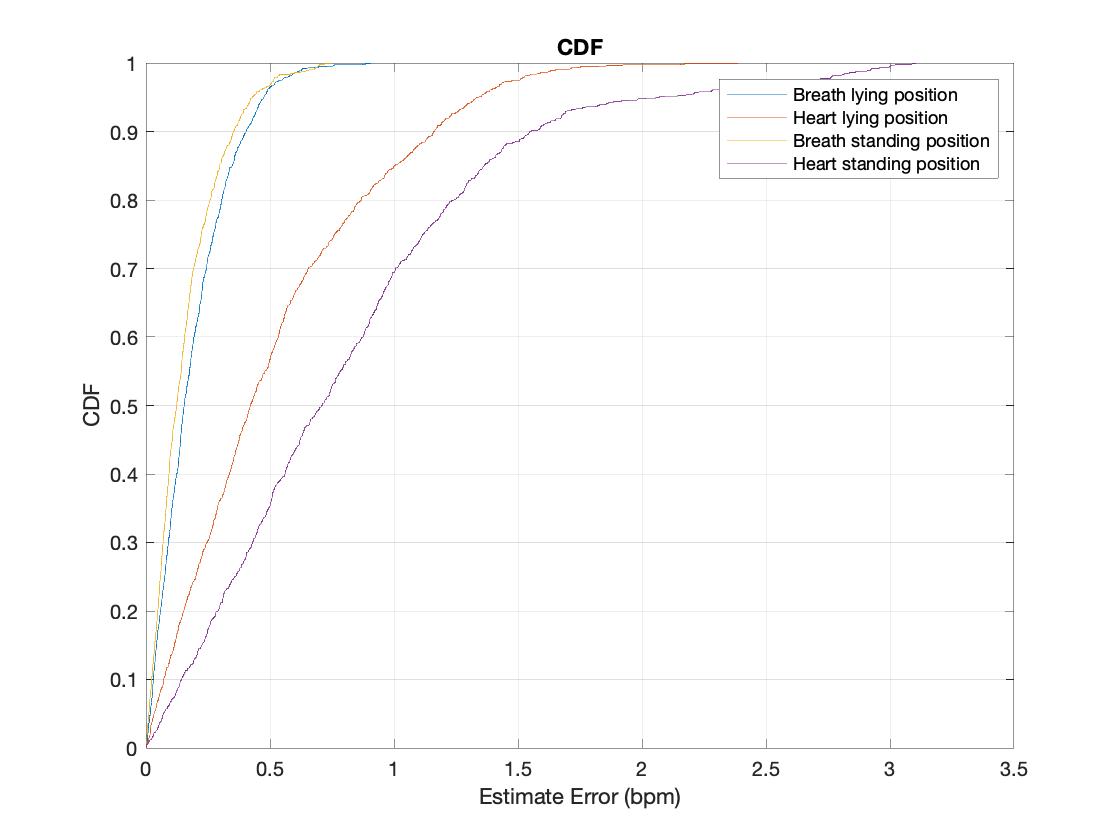}}
\caption{CDF of estimation error for standing and lying positions}
\label{Fig:CDF standing and lying}
\end{figure}

We can see from Fig.~\ref{Fig:CDF standing and lying} that the lying position results in a better performance than the standing position as it has a smaller mean value and deviation. Over 70$\%$ of respiration rate estimation errors in the lying position are less than 0.2 bpm, while over 95$\%$ of heart rate estimation errors in the lying position are less than 0.6 bpm. For the lying position, the accuracy measured by mean absolute percentage error is 99.272$\%$ for respiration rate estimation and 98.591$\%$ for the heart rate estimation. For the standing position, it is 99.1032$\%$ and 98.666$\%$, respectively. 

\subsection{Sleep Stage Classification}
\paragraph{Effect of Different Parameters} We have developed three neural networks for WiFi based sleep stage detection. One is the W2SN as described in Fig. \ref{Fig:The structure of Sleep Network}, another is a network with two convolutional layers, the third one has five convolutional layers. The structure of the two convolutional layer model is straightforward, but the performance is poor with a precision of 63.275$\%$. For the neural network with five convolutional layers, it is over-fitting. However, the accuracy for our W2SN with four convolutional layers is 95.925$\%$. In comparison, the accuracy of \cite{yu2021wifi} is 81.8$\%$ and the accuracy of \cite{deng2017decision} is 74.3$\%$. 

For the CPC neural network, we tried different structures. The classification average accuracy of two convolutional layers is 72.525$\%$. The accuracy of six convolutional layers is 71.45$\%$. However, the accuracy for our neural network (the neural network structure shown in Section \uppercase\expandafter{\romannumeral5}) is 90.15$\%$.

\paragraph{Performance of Sleep Stage Classification}

\begin{figure}
\centerline{\includegraphics[width=0.53\textwidth]{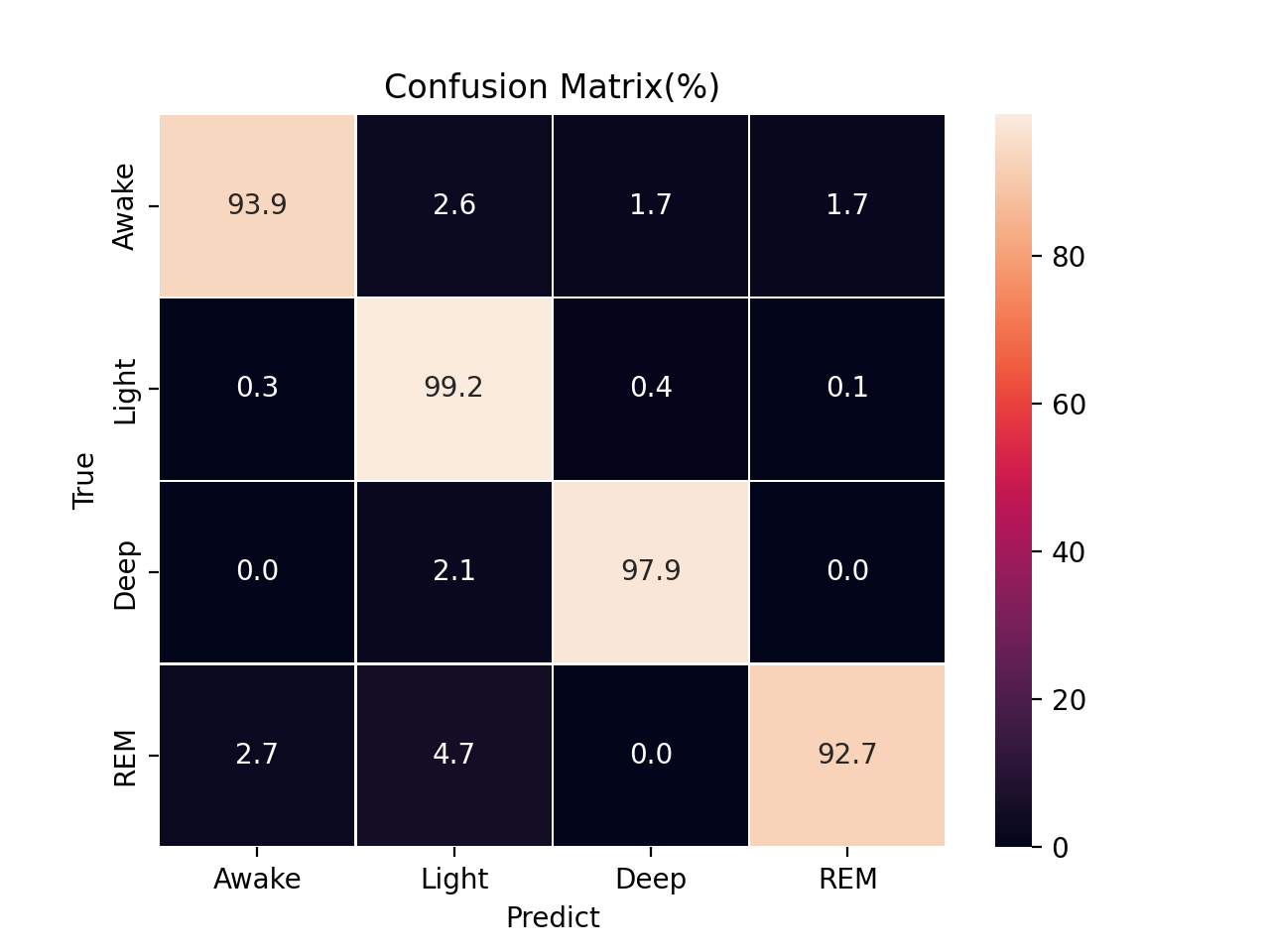}}
\caption{Confusion matrix for our system}
\label{Fig:Confusion matrix of system}
\end{figure}

Based on our proposed W2SN, the confusion matrix for sleep stage classification is shown in Fig.~\ref{Fig:Confusion matrix of system}. The actual user sleep stage is presented at each row, and each column depicts the sleep stage identified by the W2SN network. In the confusion matrix, each cell includes the actual sleep stage percentage in the row, which can be classified as the sleep stage in the column. Overall, the accuracy for each sleep stage classification is over 92$\%$. 

\begin{figure}
\centerline{\includegraphics[width=0.53\textwidth]{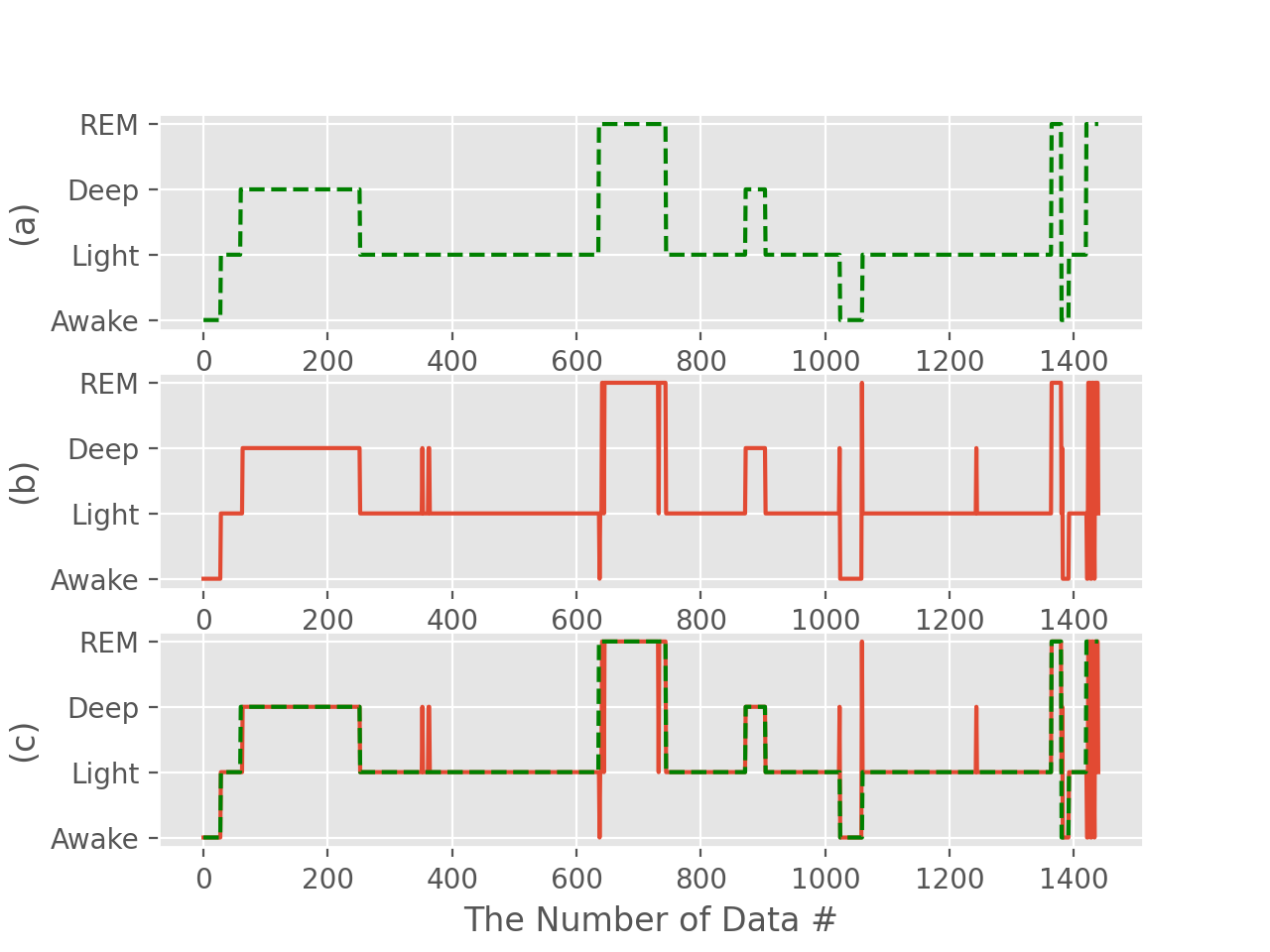}}
\caption{The sleep-stage classification results over a night}
\label{Fig:The sleep stage}
\end{figure}

The findings of sleep stage classification over a night are shown in Fig.~\ref{Fig:The sleep stage}, where the green dashed line indicates the actual sleep stage. The red line represents the estimation results. The bottom curve of Fig.~\ref{Fig:The sleep stage}  shows the error of the sleep stage classification. For wrong classifications in the graph, we can add a detector. It is possible to add an error corrector based on sleep cycle knowledge or based on the correct historical outputs to match the sleep cycle.

For the method based on CPC sleep stage neural network, the average identification accuracy for four sleep stages is 90.15$\%$. In comparison, the average accuracy of identifying sleep stages based on W2SN can reach up to 95.925$\%$.

\begin{figure}
\centerline{\includegraphics[width=0.53\textwidth]{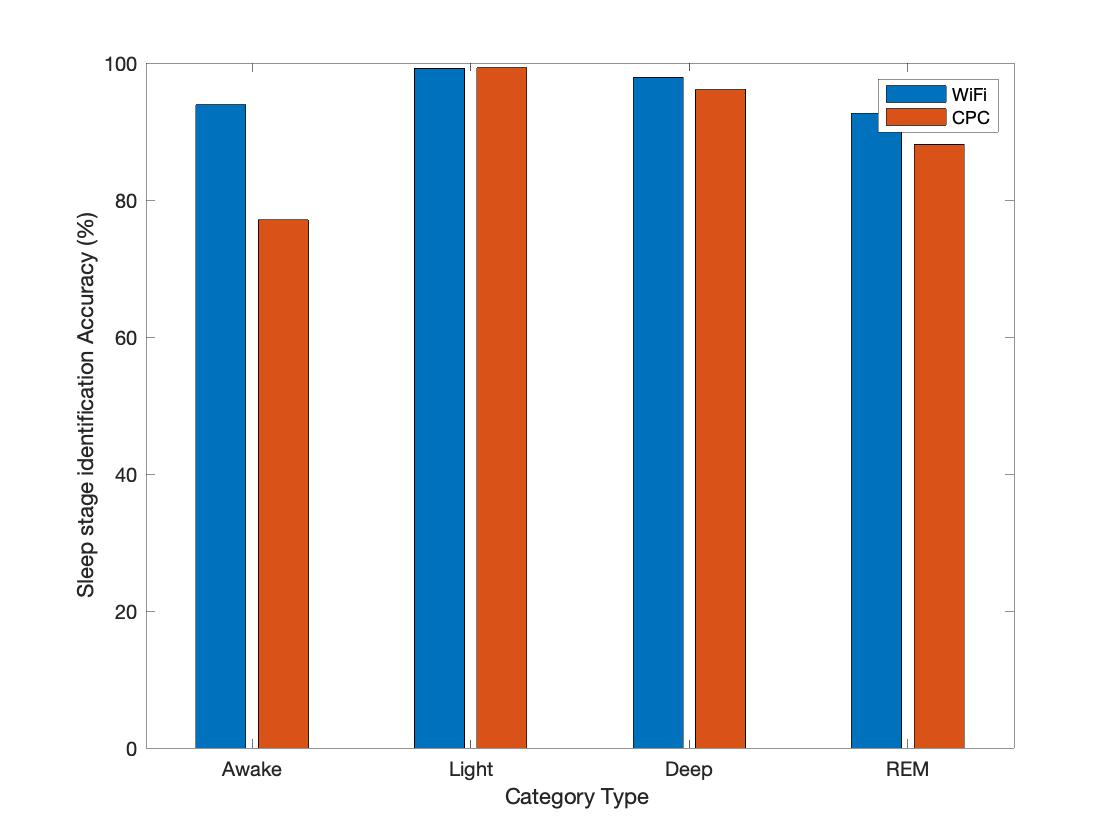}}
\caption{Performance of the sleep stage identification}
\label{Fig:compare}
\end{figure}

As shown in Fig.~\ref{Fig:compare}, the approach based on WiFi performs better than the one based on CPC. The possible reason could be that the CSI data obtained by WiFi router contains more information, i.e. human body movement during sleep, respiration rate and heart rate, however, the CPC signal only has the information for respiration rate and heart rate. Consequently, the WiFi sleep stage neural network has better performance. Since the amount of data in categories REM and Awake in our sleep database is modest, 4.7$\%$ and 12.2$\%$, respectively, as shown in Fig.~\ref{Fig:compare}, the classification performance of these two categories can be improved in the future. 

In this paper, we design two classification approaches based on deep learning. The first method uses the data which is calculated by the CPC algorithm as an input. The second method uses the unprocessed CSI matrix as an input. By comparing two neural network classification approaches, we find that the WiFi sleep stage neural network  performs better because the CSI data contains more information than the CPC signal. The accuracy of identification is 95.925$\%$. We also compared with the existing techniques in literature, e.g. \cite{yu2021wifi,deng2017decision} and found that the accuracy for sleep stage identification is improved. 

\section{Conclusions}
In this paper, we first proposed a single input multiple output CNN network to estimate both heart rate and respiration rate simultaneously. Instead of the traditional complex feature selection algorithms, it is designed with the aim of reducing the computational complexity and improving the system efficiency. In addition, we designed and compared two neural network classification approaches based on WiFi and CPC algorithms. Our system can classify four different sleep stages, including wake, light sleep, deep sleep, and REM. The estimation error for recognising respiration rate is 0.2 bpm and 0.6042 bpm for identifying the heart rate. The accuracy of the system, measured by mean absolute percentage error, is $99.109\%$ and $98.581\%$, respectively. For the lying position, the accuracy measured by the mean absolute percentage error
is $99.272\%$ for recognising respiration rate and $98.591\%$ for identifying the heart rate. For the standing position, it is $99.1032\%$ and $98.666\%$, respectively. The accuracy of the identification sleep stage is $95.925\%$. Our WiFi-based approach outperforms the state-of-the-art techniques and represents a practical and viable solution for smart health and smart medical care applications.


\bibliographystyle{IEEEtran}
\bibliography{IEEEabrv,Ref}

\begin{thebibliography}{10}
\providecommand{\url}[1]{#1}
\csname url@samestyle\endcsname
\providecommand{\newblock}{\relax}
\providecommand{\bibinfo}[2]{#2}
\providecommand{\BIBentrySTDinterwordspacing}{\spaceskip=0pt\relax}
\providecommand{\BIBentryALTinterwordstretchfactor}{4}
\providecommand{\BIBentryALTinterwordspacing}{\spaceskip=\fontdimen2\font plus
\BIBentryALTinterwordstretchfactor\fontdimen3\font minus
  \fontdimen4\font\relax}
\providecommand{\BIBforeignlanguage}[2]{{%
\expandafter\ifx\csname l@#1\endcsname\relax
\typeout{** WARNING: IEEEtran.bst: No hyphenation pattern has been}%
\typeout{** loaded for the language `#1'. Using the pattern for}%
\typeout{** the default language instead.}%
\else
\language=\csname l@#1\endcsname
\fi
#2}}
\providecommand{\BIBdecl}{\relax}
\BIBdecl

\bibitem{boric2002wireless}
O.~Boric-Lubeke and V.~M. Lubecke, ``Wireless house calls: using communications
  technology for health care and monitoring,'' \emph{IEEE Microwave Magazine},
  vol.~3, no.~3, pp. 43--48, 2002.

\bibitem{caples2005obstructive}
S.~M. Caples, A.~S. Gami, and V.~K. Somers, ``Obstructive sleep apnea,''
  \emph{Annals of internal medicine}, vol. 142, no.~3, pp. 187--197, 2005.

\bibitem{braun2012bridging}
P.~X. Braun, C.~F. Gmachl, and R.~A. Dweik, ``Bridging the collaborative gap:
  Realizing the clinical potential of breath analysis for disease diagnosis and
  monitoring--tutorial,'' \emph{IEEE Sensors Journal}, vol.~12, no.~11, pp.
  3258--3270, 2012.

\bibitem{farrell2017recognition}
P.~C. Farrell and G.~Richards, ``Recognition and treatment of sleep-disordered
  breathing: an important component of chronic disease management,''
  \emph{Journal of translational medicine}, vol.~15, no.~1, pp. 1--12, 2017.

\bibitem{tasali2008slow}
E.~Tasali, R.~Leproult, D.~A. Ehrmann, and E.~Van~Cauter, ``Slow-wave sleep and
  the risk of type 2 diabetes in humans,'' \emph{Proceedings of the National
  Academy of Sciences}, vol. 105, no.~3, pp. 1044--1049, 2008.

\bibitem{parish2009sleep}
J.~M. Parish, ``Sleep-related problems in common medical conditions,''
  \emph{Chest}, vol. 135, no.~2, pp. 563--572, 2009.

\bibitem{roth2009slow}
T.~Roth, ``Slow wave sleep: does it matter?'' \emph{Journal of Clinical Sleep
  Medicine}, vol.~5, no. 2 suppl, pp. S4--S5, 2009.

\bibitem{zeng2018fullbreathe}
Y.~Zeng, D.~Wu, R.~Gao, T.~Gu, and D.~Zhang, ``Fullbreathe: Full human
  respiration detection exploiting complementarity of csi phase and amplitude
  of wifi signals,'' \emph{Proceedings of the ACM on Interactive, Mobile,
  Wearable and Ubiquitous Technologies}, vol.~2, no.~3, pp. 1--19, 2018.

\bibitem{liu2018monitoring}
J.~Liu, Y.~Chen, Y.~Wang, X.~Chen, J.~Cheng, and J.~Yang, ``Monitoring vital
  signs and postures during sleep using wifi signals,'' \emph{IEEE Internet of
  Things Journal}, vol.~5, no.~3, pp. 2071--2084, 2018.

\bibitem{khamis2018cardiofi}
A.~Khamis, C.~T. Chou, B.~Kusy, and W.~Hu, ``Cardiofi: Enabling heart rate
  monitoring on unmodified cots wifi devices,'' in \emph{Proceedings of the
  15th EAI International Conference on Mobile and Ubiquitous Systems:
  Computing, Networking and Services}, 2018, pp. 97--106.

\bibitem{li2016mo}
F.~Li, C.~Xu, Y.~Liu, Y.~Zhang, Z.~Li, K.~Sharif, and Y.~Wang, ``Mo-sleep:
  Unobtrusive sleep and movement monitoring via wi-fi signal,'' in \emph{2016
  IEEE 35th International Performance Computing and Communications Conference
  (IPCCC)}.\hskip 1em plus 0.5em minus 0.4em\relax IEEE, 2016, pp. 1--8.

\bibitem{sleep1998methods}
S.~H. H. R. G. R.~S. sxr15@ po. cwru.~edu Sanders Mark H. Lind Bonnie K. Quan
  Stuart F. Iber Conrad Gottlieb Daniel J. Bonekat William H. Rapoport David M.
  Smith Philip L. Kiley James~P., ``Methods for obtaining and analyzing
  unattended polysomnography data for a multicenter study,'' \emph{Sleep},
  vol.~21, no.~7, pp. 759--767, 1998.

\bibitem{van2011objective}
A.~T. Van De~Water, A.~Holmes, and D.~A. Hurley, ``Objective measurements of
  sleep for non-laboratory settings as alternatives to polysomnography--a
  systematic review,'' \emph{Journal of sleep research}, vol.~20, no. 1pt2, pp.
  183--200, 2011.

\bibitem{alian2014photoplethysmography}
A.~A. Alian and K.~H. Shelley, ``Photoplethysmography,'' \emph{Best Practice \&
  Research Clinical Anaesthesiology}, vol.~28, no.~4, pp. 395--406, 2014.

\bibitem{nilsson2013respiration}
L.~M. Nilsson, ``Respiration signals from photoplethysmography,''
  \emph{Anesthesia \& Analgesia}, vol. 117, no.~4, pp. 859--865, 2013.

\bibitem{karmakar2013detection}
C.~Karmakar, A.~Khandoker, T.~Penzel, C.~Sch{\"o}bel, and M.~Palaniswami,
  ``Detection of respiratory arousals using photoplethysmography (ppg) signal
  in sleep apnea patients,'' \emph{IEEE journal of biomedical and health
  informatics}, vol.~18, no.~3, pp. 1065--1073, 2013.

\bibitem{lazaro2012osas}
J.~L{\'a}zaro, E.~Gil, J.~M. Vergara, and P.~Laguna, ``Osas detection in
  children by using ppg amplitude fluctuation decreases and pulse rate
  variability,'' in \emph{2012 Computing in Cardiology}.\hskip 1em plus 0.5em
  minus 0.4em\relax IEEE, 2012, pp. 185--188.

\bibitem{haba2005obstructive}
J.~Haba-Rubio, G.~Darbellay, F.~R. Herrmann, J.~G. Frey, A.~Fernandes, J.~M.
  Vesin, J.~P. Thiran, and J.~M. Tschopp, ``Obstructive sleep apnea syndrome:
  effect of respiratory events and arousal on pulse wave amplitude measured by
  photoplethysmography in nrem sleep,'' \emph{Sleep and Breathing}, vol.~9,
  no.~2, pp. 73--81, 2005.

\bibitem{lanata2014complexity}
A.~Lanata, G.~Valenza, M.~Nardelli, C.~Gentili, and E.~P. Scilingo,
  ``Complexity index from a personalized wearable monitoring system for
  assessing remission in mental health,'' \emph{IEEE Journal of Biomedical and
  health Informatics}, vol.~19, no.~1, pp. 132--139, 2014.

\bibitem{lopez2015wearable}
N.~Lopez-Ruiz, J.~Lopez-Torres, M.~{\'A}.~C. Rodr{\'\i}guez, I.~P.
  de~Vargas-Sansalvador, and A.~Martinez-Olmos, ``Wearable system for
  monitoring of oxygen concentration in breath based on optical sensor,''
  \emph{IEEE Sensors Journal}, vol.~15, no.~7, pp. 4039--4045, 2015.

\bibitem{sadek2018nonintrusive}
I.~Sadek and M.~Mohktari, ``Nonintrusive remote monitoring of sleep in
  home-based situation,'' \emph{Journal of medical systems}, vol.~42, no.~4,
  pp. 1--10, 2018.

\bibitem{ren2015fine}
Y.~Ren, C.~Wang, J.~Yang, and Y.~Chen, ``Fine-grained sleep monitoring: Hearing
  your breathing with smartphones,'' in \emph{2015 IEEE Conference on Computer
  Communications (INFOCOM)}.\hskip 1em plus 0.5em minus 0.4em\relax IEEE, 2015,
  pp. 1194--1202.

\bibitem{rofouei2011non}
M.~Rofouei, M.~Sinclair, R.~Bittner, T.~Blank, N.~Saw, G.~DeJean, and
  J.~Heffron, ``A non-invasive wearable neck-cuff system for real-time sleep
  monitoring,'' in \emph{2011 international conference on body sensor
  networks}.\hskip 1em plus 0.5em minus 0.4em\relax IEEE, 2011, pp. 156--161.

\bibitem{lee2016analysis}
S.~Lee, S.~Nam, and H.~Shin, ``The analysis of sleep stages with motion and
  heart rate signals from a handheld wearable device,'' in \emph{2016
  International Conference on Information and Communication Technology
  Convergence (ICTC)}.\hskip 1em plus 0.5em minus 0.4em\relax IEEE, 2016, pp.
  1135--1137.

\bibitem{li2016noncontact}
M.~H. Li, A.~Yadollahi, and B.~Taati, ``Noncontact vision-based cardiopulmonary
  monitoring in different sleeping positions,'' \emph{IEEE journal of
  biomedical and health informatics}, vol.~21, no.~5, pp. 1367--1375, 2016.

\bibitem{scully2011physiological}
C.~G. Scully, J.~Lee, J.~Meyer, A.~M. Gorbach, D.~Granquist-Fraser,
  Y.~Mendelson, and K.~H. Chon, ``Physiological parameter monitoring from
  optical recordings with a mobile phone,'' \emph{IEEE Transactions on
  Biomedical Engineering}, vol.~59, no.~2, pp. 303--306, 2011.

\bibitem{boccanfuso2012remote}
L.~Boccanfuso and J.~M. O'Kane, ``Remote measurement of breathing rate in real
  time using a high precision, single-point infrared temperature sensor,'' in
  \emph{2012 4th IEEE RAS \& EMBS International Conference on Biomedical
  Robotics and Biomechatronics (BioRob)}.\hskip 1em plus 0.5em minus
  0.4em\relax IEEE, 2012, pp. 1704--1709.

\bibitem{pereira2015remote}
C.~B. Pereira, X.~Yu, M.~Czaplik, R.~Rossaint, V.~Blazek, and S.~Leonhardt,
  ``Remote monitoring of breathing dynamics using infrared thermography,''
  \emph{Biomedical optics express}, vol.~6, no.~11, pp. 4378--4394, 2015.

\bibitem{murthy2006noncontact}
R.~Murthy and I.~Pavlidis, ``Noncontact measurement of breathing function,''
  \emph{IEEE Engineering in medicine and biology magazine}, vol.~25, no.~3, pp.
  57--67, 2006.

\bibitem{li2009accurate}
C.~Li, J.~Ling, J.~Li, and J.~Lin, ``Accurate doppler radar noncontact vital
  sign detection using the relax algorithm,'' \emph{IEEE Transactions on
  Instrumentation and Measurement}, vol.~59, no.~3, pp. 687--695, 2009.

\bibitem{salmi2011propagation}
J.~Salmi and A.~F. Molisch, ``Propagation parameter estimation, modeling and
  measurements for ultrawideband mimo radar,'' \emph{IEEE Transactions on
  Antennas and Propagation}, vol.~59, no.~11, pp. 4257--4267, 2011.

\bibitem{nguyen2016continuous}
P.~Nguyen, X.~Zhang, A.~Halbower, and T.~Vu, ``Continuous and fine-grained
  breathing volume monitoring from afar using wireless signals,'' in \emph{IEEE
  INFOCOM 2016-The 35th Annual IEEE International Conference on Computer
  Communications}.\hskip 1em plus 0.5em minus 0.4em\relax IEEE, 2016, pp. 1--9.

\bibitem{park2006single}
B.-K. Park, S.~Yamada, O.~Boric-Lubecke, and V.~Lubecke, ``Single-channel
  receiver limitations in doppler radar measurements of periodic motion,'' in
  \emph{2006 IEEE Radio and Wireless Symposium}.\hskip 1em plus 0.5em minus
  0.4em\relax IEEE, 2006, pp. 99--102.

\bibitem{cianca2009fm}
E.~Cianca and B.~Gupta, ``Fm-uwb for communications and radar in medical
  applications,'' \emph{Wireless Personal Communications}, vol.~51, no.~4, p.
  793, 2009.

\bibitem{adib2015smart}
F.~Adib, H.~Mao, Z.~Kabelac, D.~Katabi, and R.~C. Miller, ``Smart homes that
  monitor breathing and heart rate,'' in \emph{Proceedings of the 33rd annual
  ACM conference on human factors in computing systems}, 2015, pp. 837--846.

\bibitem{GI2018xiaopeng}
X.~Wang and Z.~Lin, ``Nonrandom microwave ghost imaging,'' \emph{IEEE
  Transactions on Geoscience and Remote Sensing}, vol.~56, no.~8, pp.
  4747--4764, 2018.

\bibitem{2016xiaopeng}
------, ``Microwave surveillance based on ghost imaging and distributed
  antennas,'' \emph{IEEE Antennas and Wireless Propagation Letters}, vol.~15,
  pp. 1831--1834, 2016.

\bibitem{ziqian2018}
Z.~Zhang, R.~Luo, X.~Wang, and Z.~Lin, ``Microwave ghost imaging via lte-dl
  signals,'' in \emph{2018 International Conference on Radar (RADAR)}, 2018,
  pp. 1--5.

\bibitem{zhao2017learning}
M.~Zhao, S.~Yue, D.~Katabi, T.~S. Jaakkola, and M.~T. Bianchi, ``Learning sleep
  stages from radio signals: A conditional adversarial architecture,'' in
  \emph{International Conference on Machine Learning}.\hskip 1em plus 0.5em
  minus 0.4em\relax PMLR, 2017, pp. 4100--4109.

\bibitem{pittella2017cardiorespiratory}
E.~Pittella, A.~Bottiglieri, S.~Pisa, and M.~Cavagnaro, ``Cardiorespiratory
  frequency monitoring using the principal component analysis technique on uwb
  radar signal,'' \emph{International Journal of Antennas and Propagation},
  vol. 2017, 2017.

\bibitem{cho2017novel}
H.-S. Cho, Y.-J. Park, H.-K. Lyu, and J.-H. Cho, ``Novel heart rate detection
  method using uwb impulse radar,'' \emph{Journal of Signal Processing
  Systems}, vol.~87, no.~2, pp. 229--239, 2017.

\bibitem{wu2012fila}
K.~Wu, J.~Xiao, Y.~Yi, M.~Gao, and L.~M. Ni, ``Fila: Fine-grained indoor
  localization,'' in \emph{2012 Proceedings IEEE INFOCOM}.\hskip 1em plus 0.5em
  minus 0.4em\relax IEEE, 2012, pp. 2210--2218.

\bibitem{wang2014eyes}
Y.~Wang, J.~Liu, Y.~Chen, M.~Gruteser, J.~Yang, and H.~Liu, ``E-eyes:
  device-free location-oriented activity identification using fine-grained wifi
  signatures,'' in \emph{Proceedings of the 20th annual international
  conference on Mobile computing and networking}, 2014, pp. 617--628.

\bibitem{zhou2015wifi}
Z.~Zhou, Z.~Yang, C.~Wu, L.~Shangguan, H.~Cai, Y.~Liu, and L.~M. Ni,
  ``Wifi-based indoor line-of-sight identification,'' \emph{IEEE Transactions
  on Wireless Communications}, vol.~14, no.~11, pp. 6125--6136, 2015.

\bibitem{yang2013rssi}
Z.~Yang, Z.~Zhou, and Y.~Liu, ``From rssi to csi: Indoor localization via
  channel response,'' \emph{ACM Computing Surveys (CSUR)}, vol.~46, no.~2, pp.
  1--32, 2013.

\bibitem{ruichen2018}
R.~Luo, Z.~Zhang, X.~Wang, and Z.~Lin, ``Wi-fi based device-free microwave
  ghost imaging indoor surveillance system,'' in \emph{2018 28th International
  Telecommunication Networks and Applications Conference (ITNAC)}, 2018, pp.
  1--6.

\bibitem{wang2015understanding}
W.~Wang, A.~X. Liu, M.~Shahzad, K.~Ling, and S.~Lu, ``Understanding and
  modeling of wifi signal based human activity recognition,'' in
  \emph{Proceedings of the 21st annual international conference on mobile
  computing and networking}, 2015, pp. 65--76.

\bibitem{ohara2017detecting}
K.~Ohara, T.~Maekawa, and Y.~Matsushita, ``Detecting state changes of indoor
  everyday objects using wi-fi channel state information,'' \emph{Proceedings
  of the ACM on interactive, mobile, wearable and ubiquitous technologies},
  vol.~1, no.~3, pp. 1--28, 2017.

\bibitem{wang2016wifall}
Y.~Wang, K.~Wu, and L.~M. Ni, ``Wifall: Device-free fall detection by wireless
  networks,'' \emph{IEEE Transactions on Mobile Computing}, vol.~16, no.~2, pp.
  581--594, 2016.

\bibitem{DiZhai2015}
D.~Zhai and Z.~Lin, ``Rss-based indoor positioning with biased estimator and
  local geographical factor,'' in \emph{2015 22nd International Conference on
  Telecommunications (ICT)}, 2015, pp. 398--402.

\bibitem{patwari2013breathfinding}
N.~Patwari, L.~Brewer, Q.~Tate, O.~Kaltiokallio, and M.~Bocca, ``Breathfinding:
  A wireless network that monitors and locates breathing in a home,''
  \emph{IEEE Journal of Selected Topics in Signal Processing}, vol.~8, no.~1,
  pp. 30--42, 2013.

\bibitem{abdelnasser2015ubibreathe}
H.~Abdelnasser, K.~A. Harras, and M.~Youssef, ``Ubibreathe: A ubiquitous
  non-invasive wifi-based breathing estimator,'' in \emph{Proceedings of the
  16th ACM International Symposium on Mobile Ad Hoc Networking and Computing},
  2015, pp. 277--286.

\bibitem{patwari2013monitoring}
N.~Patwari, J.~Wilson, S.~Ananthanarayanan, S.~K. Kasera, and D.~R. Westenskow,
  ``Monitoring breathing via signal strength in wireless networks,'' \emph{IEEE
  Transactions on Mobile Computing}, vol.~13, no.~8, pp. 1774--1786, 2013.

\bibitem{zhang2019breathtrack}
D.~Zhang, Y.~Hu, Y.~Chen, and B.~Zeng, ``Breathtrack: Tracking indoor human
  breath status via commodity wifi,'' \emph{IEEE Internet of Things Journal},
  vol.~6, no.~2, pp. 3899--3911, 2019.

\bibitem{yu2021wifi}
B.~Yu, Y.~Wang, K.~Niu, Y.~Zeng, T.~Gu, L.~Wang, C.~Guan, and D.~Zhang,
  ``Wifi-sleep: Sleep stage monitoring using commodity wi-fi devices,''
  \emph{IEEE Internet of Things Journal}, 2021.

\bibitem{lee2018design}
S.~Lee, Y.-D. Park, Y.-J. Suh, and S.~Jeon, ``Design and implementation of
  monitoring system for breathing and heart rate pattern using wifi signals,''
  in \emph{2018 15th IEEE Annual Consumer Communications \& Networking
  Conference (CCNC)}.\hskip 1em plus 0.5em minus 0.4em\relax IEEE, 2018, pp.
  1--7.

\bibitem{hussain2021vehicle}
M.~Hussain, A.~Akbilek, F.~Pfeiffer, and B.~Napholz, ``In-vehicle breathing
  rate monitoring based on wifi signals,'' in \emph{2020 50th European
  Microwave Conference (EuMC)}.\hskip 1em plus 0.5em minus 0.4em\relax IEEE,
  2021, pp. 292--295.

\bibitem{deng2017decision}
B.~Deng, B.~Xue, H.~Hong, C.~Fu, X.~Zhu, and Z.~Wang, ``Decision tree based
  sleep stage estimation from nocturnal audio signals,'' in \emph{2017 22nd
  International Conference on Digital Signal Processing (DSP)}.\hskip 1em plus
  0.5em minus 0.4em\relax IEEE, 2017, pp. 1--4.

\bibitem{dafna2016estimation}
E.~Dafna, M.~Halevi, D.~B. Or, A.~Tarasiuk, and Y.~Zigel, ``Estimation of macro
  sleep stages from whole night audio analysis,'' in \emph{2016 38th Annual
  International Conference of the IEEE Engineering in Medicine and Biology
  Society (EMBC)}.\hskip 1em plus 0.5em minus 0.4em\relax IEEE, 2016, pp.
  2847--2850.

\bibitem{dafna2015sleep}
E.~Dafna, A.~Tarasiuk, and Y.~Zigel, ``Sleep-wake evaluation from whole-night
  non-contact audio recordings of breathing sounds,'' \emph{PloS one}, vol.~10,
  no.~2, p. e0117382, 2015.

\bibitem{nandakumar2015contactless}
R.~Nandakumar, S.~Gollakota, and N.~Watson, ``Contactless sleep apnea detection
  on smartphones,'' in \emph{Proceedings of the 13th annual international
  conference on mobile systems, applications, and services}, 2015, pp. 45--57.

\bibitem{gu2019wifi}
Y.~Gu, X.~Zhang, Z.~Liu, and F.~Ren, ``Wifi-based real-time breathing and heart
  rate monitoring during sleep,'' in \emph{2019 IEEE Global Communications
  Conference (GLOBECOM)}.\hskip 1em plus 0.5em minus 0.4em\relax IEEE, 2019,
  pp. 1--6.

\bibitem{wang2016human}
H.~Wang, D.~Zhang, J.~Ma, Y.~Wang, Y.~Wang, D.~Wu, T.~Gu, and B.~Xie, ``Human
  respiration detection with commodity wifi devices: do user location and body
  orientation matter?'' in \emph{Proceedings of the 2016 ACM International
  Joint Conference on Pervasive and Ubiquitous Computing}, 2016, pp. 25--36.

\bibitem{wu2017device}
D.~Wu, D.~Zhang, C.~Xu, H.~Wang, and X.~Li, ``Device-free wifi human sensing:
  From pattern-based to model-based approaches,'' \emph{IEEE Communications
  Magazine}, vol.~55, no.~10, pp. 91--97, 2017.

\bibitem{web}
``Sydney iot platform, electrocardiogram (ecg) monitoring module,''
  \url{https://www.sydneyiot.com/ecg$-$monitoring$-$module}, accessed January
  11, 2022.

\bibitem{xucun2020WCNC}
X.~Yan, Z.~Lin, and P.~Wang, ``Wireless electrocardiograph monitoring based on
  wavelet convolutional neural network,'' in \emph{2020 IEEE Wireless
  Communications and Networking Conference Workshops (WCNCW)}, 2020, pp. 1--6.

\bibitem{menglu2021}
L.~Meng, K.~Ge, Y.~Song, D.~Yang, and Z.~Lin, ``Long-term wearable
  electrocardiogram signal monitoring and analysis based on convolutional
  neural network,'' \emph{IEEE Transactions on Instrumentation and
  Measurement}, vol.~70, pp. 1--11, 2021.

\bibitem{zijiao2021}
Z.~Chen, Z.~Lin, P.~Wang, and M.~Ding, ``Negative-resnet: noisy ambulatory
  electrocardiogram signal classification scheme,'' \emph{Neural Computing and
  Applications}, vol.~33, pp. 1--13, 07 2021.

\bibitem{wang2022wearable}
P.~Wang, Z.~Lin, X.~Yan, Z.~Chen, M.~Ding, Y.~Song, and L.~Meng, ``A wearable
  ecg monitor for deep learning based real-time cardiovascular disease
  detection,'' \emph{arXiv preprint arXiv:2201.10083}, 2022.

\bibitem{elgendi2013fast}
M.~Elgendi, ``Fast qrs detection with an optimized knowledge-based method:
  Evaluation on 11 standard ecg databases,'' \emph{PloS one}, vol.~8, no.~9, p.
  e73557, 2013.

\bibitem{lipsitz1995heart}
L.~A. Lipsitz, F.~Hashimoto, L.~P. Lubowsky, J.~Mietus, G.~B. Moody,
  O.~Appenzeller, and A.~L. Goldberger, ``Heart rate and respiratory rhythm
  dynamics on ascent to high altitude.'' \emph{Heart}, vol.~74, no.~4, pp.
  390--396, 1995.

\bibitem{wallin2010relationship}
B.~Wallin, E.~Hart, E.~A. Wehrwein, N.~Charkoudian, and M.~Joyner,
  ``Relationship between breathing and cardiovascular function at rest:
  sex-related differences,'' \emph{Acta physiologica}, vol. 200, no.~2, pp.
  193--200, 2010.

\bibitem{berry2012aasm}
R.~B. Berry, R.~Brooks, C.~E. Gamaldo, S.~M. Harding, C.~Marcus, B.~V. Vaughn
  \emph{et~al.}, ``The aasm manual for the scoring of sleep and associated
  events,'' \emph{Rules, Terminology and Technical Specifications, Darien,
  Illinois, American Academy of Sleep Medicine}, vol. 176, p. 2012, 2012.

\bibitem{lin2016sleepsense}
F.~Lin, Y.~Zhuang, C.~Song, A.~Wang, Y.~Li, C.~Gu, C.~Li, and W.~Xu,
  ``Sleepsense: A noncontact and cost-effective sleep monitoring system,''
  \emph{IEEE transactions on biomedical circuits and systems}, vol.~11, no.~1,
  pp. 189--202, 2016.

\bibitem{wilde1983rate}
J.~Wilde-Frenz and H.~Schulz, ``Rate and distribution of body movements during
  sleep in humans,'' \emph{Perceptual and motor skills}, vol.~56, no.~1, pp.
  275--283, 1983.

\bibitem{chen2018deepphys}
W.~Chen and D.~McDuff, ``Deepphys: Video-based physiological measurement using
  convolutional attention networks,'' in \emph{Proceedings of the European
  Conference on Computer Vision (ECCV)}, 2018, pp. 349--365.

\bibitem{yu2019remote}
Z.~Yu, X.~Li, and G.~Zhao, ``Remote photoplethysmograph signal measurement from
  facial videos using spatio-temporal networks,'' \emph{arXiv preprint
  arXiv:1905.02419}, 2019.

\bibitem{lee2020meta}
E.~Lee, E.~Chen, and C.-Y. Lee, ``Meta-rppg: Remote heart rate estimation using
  a transductive meta-learner,'' in \emph{European Conference on Computer
  Vision}.\hskip 1em plus 0.5em minus 0.4em\relax Springer, 2020, pp. 392--409.

\bibitem{ni2021review}
A.~Ni, A.~Azarang, and N.~Kehtarnavaz, ``A review of deep learning-based
  contactless heart rate measurement methods,'' \emph{Sensors}, vol.~21,
  no.~11, p. 3719, 2021.

\end{thebibliography}
\end{document}